\documentclass{article}
\usepackage{arxiv}

\usepackage[utf8]{inputenc}
\usepackage{textcomp}
\usepackage{microtype}
\usepackage{graphicx}
\usepackage{booktabs} % for professional tables
\usepackage{hyperref}

\usepackage[utf8]{inputenc}
\usepackage{xcolor}
\usepackage{fancybox}
\usepackage{graphicx}
\usepackage{amsmath}
\usepackage{amsfonts}
\usepackage{amssymb}
\usepackage{gensymb}
\usepackage{mathtools}
\usepackage{eurosym, booktabs, units}
\usepackage{pgfgantt}
\usepackage{enumitem}
\usepackage{colortbl}
\usepackage{tabularx}
\usepackage{caption}
\usepackage{subcaption}
\usepackage{float}
\usepackage{algorithmicx}
\usepackage{algorithm}
\usepackage{algpseudocode}
\usepackage{siunitx}
\usepackage{ bbold }
\usepackage{nicefrac}       % compact symbols for 1/2, etc.
\usepackage{amsthm}

\theoremstyle{definition}
\newtheorem{definition}{Definition}[section]
\theoremstyle{remark}
\newtheorem*{remark}{Remark}
\newtheorem{assumption}{Assumption}
\newtheorem{theorem}{Theorem}
\newtheorem{lemma}{Lemma}

\newcommand{\displaycomments}
\ifdefined\displaycomments

\newif\ifAddInBlue
\AddInBluetrue   % comment out for print

\newif\ifShowDel
\newcommand{\del}[1]{%
  \ifShowDel
  \textcolor{red}{#1}	
  \fi%
}

\title{Learning Constrained Adaptive Differentiable Predictive Control Policies With Guarantees}

\author{J\'an Drgo\v na, Aaron Tuor, Draguna Vrabie \\
    Pacific Northwest National Laboratory\\
	Richland, Washington USA\\
	\{jan.drgona, aaron.tuor, draguna.vrabie\}@pnnl.gov
}

\begin{document}

\maketitle

\begin{abstract}
We present differentiable predictive control (DPC), a  method for learning constrained neural control policies for linear systems with probabilistic performance guarantees. 
We employ automatic differentiation to obtain direct policy gradients by backpropagating the model predictive control (MPC) loss function and constraints penalties through a differentiable closed-loop system dynamics model.
We demonstrate that the proposed method can learn parametric constrained control policies to stabilize systems with unstable dynamics, track time-varying references, and satisfy nonlinear state and input constraints. In contrast with imitation learning-based approaches, our method does not depend on a supervisory controller.
Most importantly, we demonstrate that, without losing performance, our method is scalable and computationally more efficient than implicit, explicit, and approximate MPC.

\end{abstract}

% \begin{abstract}
% We present neural differentiable predictive control (DPC)  method for learning constrained neural control policies for uncertain linear systems. 
% DPC is formulated as a differentiable constrained optimization problem whose computational graph architecture is inspired by classical model predictive control (MPC) structure.
% In particular, the optimization of the neural control policy is based on automatic differentiation of the MPC loss function and constraints penalties
% through a differentiable closed-loop system dynamics model.
% We show that DPC can learn constrained neural control policies to stabilize systems with unstable dynamics, track time-varying references, and satisfy state and input constraints without the prior need of a supervisory MPC controller.
% Most importantly, we demonstrate that DPC scales linearly with problem size, compared to exponential scalability of classical explicit MPC based on multiparametric programming.
% \end{abstract}

\section{Introduction}

% \add{mention paper: DeepMPC: Learning Deep Latent Features for Model Predictive Control - similar idea learning neural model of the dynamics and then backpropagating to get the loss function gradients for computing the optimal control actions online as in the case of online MPC, in our case we do the optimization offline }

% Many real-world systems of critical interest are characterized by multi-scale and heterogeneous nonlinear dynamics,
Many real-world systems of critical interest have unknown dynamics,
uncertain and dynamic operating environments, and 
constrained operating regimes.
This presents challenges to design efficient and robust control algorithms. Advanced control design requires expertise in applied mathematics, dynamic systems theory, computational methods, modeling, optimization, and operator assisted algorithmic tuning of control parameters.
These requirements increase  cost and reduce their applicability only to high value systems where marginal  performance improvements lead to huge economic benefits. 

Data-driven dynamics modeling and control policy learning show promise to ``democratize" advanced control to systems with complex and partially characterized dynamics. However, pure data-driven methods typically suffer from poor-sampling efficiency, scale poorly with problem size, and exhibit slow convergence to optimal decisions~\cite{RL_challenges2019}.   Of special concern  are the lack of guarantees that black-box data-driven controllers will satisfy operational constraints and maintain safe operation.

% On the other hand, model-based methods such as 
Model Predictive Control (MPC) methods can offer optimal performance with systematic constraints handling. However, their solution is based on real-time optimization that might be computationally prohibitive for some applications. 
Explicit MPC~\cite{Alessio2009} aims to overcome the limits of online optimization by pre-computing the control law offline for a given set of admissible parameters.
However, obtaining explicit MPC policies via multiparametric programming~\cite{kvasnica:vdm} is computationally expensive, limiting its applicability to small-scale systems.
Addressing limitations of explicit MPC, several authors have proposed using deep learning to approximate the control policies given example data sets computed by a supervisory MPC~\cite{maddalena2019neural,KARG2021107266}.

\subsection{Contributions}

In this work, we present a method for learning explicit model-based predictive control policies by using differentiable programming~\cite{DiffProg2019}. We call the presented method differentiable predictive control (DPC). In DPC, the mathematical operations associated with evaluating the MPC problem's constraints and objective function are represented as a directed acyclic computational graph implemented in a language supporting automatic differentiation (AD). We show that with DPC, it is possible to learn constrained neural control policies by differentiating the constraints and objectives of the MPC problem in an unsupervised way. By doing so, we alleviate the need for the solution of the original problem using an online or parametric programming solver.
In DPC, we optimize an explicit neural controller offline via stochastic gradient descent updates using direct policy gradients computed over a synthetically sampled distribution of the control parameters. The direct policy gradients are obtained via automatic differentiation of the MPC problem's constraints violations and objectives over a closed-loop system model rollouts for the given prediction horizon. 
% In particular, we compute direct policy gradients by automatic differentiation of the computational graph that corresponds to the unrolled computations of the MPC problem’s constraints and objectives
We report the following  contributions:
\begin{enumerate}
    \item A method for training differentiable computational graphs of model predictive controllers with linear state space dynamics and neural control policies, subject to nonlinear state and control action constraints.
    \item An offline unsupervised policy optimization algorithm based on policy gradients computed via automatic differentiation of the MPC problem's computational graph.
     \item An algorithm to provide probabilistic closed-loop stability and constraints satisfaction guarantees of the DPC policy, based on Hoeffding’s inequality.
     \item Sufficient conditions of closed-loop  stability for linear systems controlled by neural network policies based on contraction conditions and Banach fixed point theorem.
    \item Five numerical studies that compare DPC with LQR, MPC, approximate MPC, and reinforcement learning (RL) algorithms demonstrating that:
    \begin{itemize}
     \item DPC can learn stabilizing policies for unstable systems.
    \item DPC can systematically handle parametric nonlinear constraints and parametric references.
    \item  DPC can handle system uncertainties online via  adaptive system model and policy updates. 
    \item DPC is more sample efficient than model-free RL algorithms.
    \item DPC policies train faster than approximate MPC based on imitation learning of supervisory MPC.
    \item DPC has faster execution time than implicit MPC  based on online constrained optimization solvers.
    \item  DPC has better scalability and requires less memory in comparison with explicit MPC based on multi-parametric programming solvers.
    \end{itemize}
        \item Open-source implementation of the proposed method~\cite{Neuromancer2021}.
\end{enumerate}

\subsection{Related Work}

\subsubsection{Learning-based model predictive control}
In general, Learning-based MPC (LBMPC) methods~\cite{ASWANI20131216} are based on learning the system dynamics model from data, and can be considered  generalizations of classical adaptive MPC~\cite{aswani2014practical}.
To make LBMPC tractable, the performance and safety tasks are  decoupled by using reachability analysis~\cite{reachPWA2000,Rakovic2006}.
Variations include formulation of robust MPC with state-dependent uncertainty for data-driven linear models~\cite{SOLOPERTO2018442},
or iterative model updates for linear systems with bounded  uncertainties and robustness guarantees~\cite{Bujarbaruah2018AdaptiveMF}.
For a comprehensive review of LBMPC approaches we refer the reader to a recent review~\cite{Hewing2020} and references therein. 
  A conceptual idea of backpropagating through the learned system model parametrized via convex neural networks~\cite{AmosXK16} was investigated in~\cite{chen2018optimal}.
  Authors in~\cite{Lenz2015DeepMPCLD} introduced a new recurrent neural model for learning latent dynamics for MPC.
  However, LBMPC approaches require sets of solutions, computed online, of the corresponding optimization problem to learn from. In contrast, in the DPC methodology proposed here, the neural policy is learned offline and hence represents an explicit solution of the underlying parametric optimal control problem.

% optimal control
\subsubsection{Explicit model predictive control}
\label{sec:eMPC1}
For a certain class of small scale MPC problems~\cite{Binder2001,Maciejowski:2002}, the solution can be pre-computed offline using multiparametric programming (mpP)~\cite{BempMoraDuaPist:2000,kvasnica:vdm} to obtain a so-called
explicit  MPC (eMPC) policy~\cite{BBM00e,TonEtal:aut:01}.
The benefits of eMPC are typically faster online computations, exact bounds on worst-case execution time, and simple and verifiable policy code, which makes it a suitable approach for embedded applications.
However, eMPC  suffers from the curse of dimensionality, scaling geometrically with the number of constraints. This severely limits its practical applicability only to small scale systems with short prediction horizons~\cite{Alessio2009}, even after applying various complexity reduction methods~\cite{Kvasnica_2012,KVASNICA20131776,HOVLAND20087711,Gulan2020}.
  The DPC method proposed here presents a scalable alternative for obtaining explicit MPC policies for linear systems by employing principles of data-driven constrained differentiable programming.

\subsubsection{Approximate model predictive control}
% imitation learning
Targeting the scalability issues of explicit MPC, authors in the control community proposed approximate MPC~\cite{Domahidi2011, ZhangKLA15,DRGONA2018, RahaCRB20,chen2019large}
whose solution is based on supervised learning of  control policies imitating the original MPC. An interesting theoretical connection was made by authors in~\cite{karg2018efficient} showing that every piecewise affine (PWA) control policy can be exactly represented by a deep ReLU network~\cite{montufar2014number}. This means that the optimality of the neural control policy will depend only on the quality of training data and formulation of the learning problem. However, the  remaining disadvantage of approximate MPC is its inherent dependency on the solution of the original MPC problem. This paper presents an alternative method for computing scalable explicit control policies for linear systems subject to nonlinear constraints. Our approach is based on differentiable programming and avoids the need for a supervisory MPC controller, as in the case of approximate MPC.

Approximate MPC methods alone do not provide performance guarantees, as in the case of implicit MPC. This limitation has previously been addressed either by involving optimality checks with backup controllers~\cite{Zhang2019SafeAN}, projections of the control actions onto the feasible set~\cite{Chen2018}, or by providing sampling-based probabilistic performance guarantees~\cite{Hertneck8371312}.
In this work, we adapt probabilistic performance guarantees as introduced in~\cite{Hertneck8371312}
in the context of unsupervised learning of the constrained control policies via the proposed DPC method.

\subsubsection{Constrained deep learning}
There are numerous challenges associated with solving constrained  deep learning problems, including guarantees on
 convergence to stationary points and global minima, smart learning rates, min-max optimization, non-convex regularizers, or constraint satisfaction guarantees~\cite{ConstrainedML2019}.
% hard constraints
In generic deep learning literature, specific neural architectures can be designed to impose  a certain class of hard constraints, such as linear operator constraints~\cite{hendriks2020linearly}.
% barriers
 Authors in~\cite{logbarrierCNN2019} demonstrated that using a log-barrier method for imposing inequality constraints could lead to improved accuracy, constraint satisfaction, and training stability.
% penalties
Penalty methods based on regularization terms in the loss function have become a popular choice for imposing inequality constraints on the outputs of deep neural network models~\cite{PathakKD15,ConstrCNN7971941}. 
% soft constraints practical advantages
As pointed out by~\cite{MarquezNeilaSF17}, in practice, the existing methods for incorporating hard constraints rarely have better prediction performance than their soft constraint counterparts, despite providing weak performance guarantees. 

\subsubsection{Safe deep learning-based control}
For imposing stability guarantees in deep learning-based control applications, one could employ  data-driven methods based on learning Lyapunov function candidates 
for stable dynamics models and control policies
as part of the neural architecture~\cite{NIPS2019_9292,Donti_control2020}.
Others~\cite{Mahyar2022} have employed semidefinite programming for safety verification and robustness analysis of feed-forward neural network control policies.
Authors in~\cite{DiffCVxLayers2019} represent constrained optimization problems as implicit layers in deep neural networks, leading to the introduction of neural network policy architectures designed to handle pre-defined constraints~\cite{MPC_PInet2018,diffMPC2018,GNURL2019}.
Authors in~\cite{Zanon2019,ZanonRLMPC2019,Gros2021} provide safety guarantees in reinforcement learning (RL) by considering differentiable MPC problem as a policy approximation. 
While both~\cite{Zanon2019} and the presented DPC approach are based on differentiating the MPC problem, in DPC the computed policy gradients are used for offline optimization of an explicit neural policy rather than for obtaining safety guarantees with RL via online optimization.

\section{Background}

\subsection{Implicit Model Predictive Control} \label{sec:mpc}

Model predictive control (MPC) is an optimal control strategy that calculates the
 control inputs trajectories by minimizing a given objective function over a finite 
 prediction horizon with respect to the constraints on the system  dynamics.
We consider a reference tracking formulation of linear MPC, given as the following constrained optimization problem:
\begin{subequations}
\label{eq:mpc_example}
    \begin{align}
 \min_{{{\bf u}_0, \ldots, {\bf u}_{N-1}}} & \sum_{k=0}^{N-1} 
 \ell_{\texttt{MPC}}( {\bf x}_k, {\bf u}_k, {\bf r}_k )
 & 
 \label{eq:mpc_example:Q} \\ 
  \text{s.t.} \ &   {\bf x}_{k+1} = {\bf A} {\bf x}_k + {\bf B } {\bf u}_k, \  k \in \mathbb{N}_{0}^{N-1}   \label{eq:mpc_example:x}  & \\
  \ &  g({\bf x}_k) \leq {\bf 0}, \ k \in \mathbb{N}_{0}^{N-1}   \label{eq:mpc_example:xb} \\
    \ &  h({\bf u}_k) \leq {\bf 0}, \ k \in \mathbb{N}_{0}^{N-1}   \label{eq:mpc_example:ub} \\
 \ & {{\bf x}_0} = {\bf x}(t),  & 
\end{align}
\end{subequations}
where ${\bf x}_k \in \mathbb{R}^{n_x}$ is the system state, ${\bf u_k} \in \mathbb{R}^{n_u}$ is the control input at time $k$, and $\mathbb{N}_a^b = \{a, a+1, \ldots, b \}$ is a set of integers.
The objective~\eqref{eq:mpc_example:Q} is a differentiable function representing the control performance metric such as the following  reference tracking with control action minimization in the quadratic form:
\begin{equation}
\label{eq:MPC_obj}
     \ell_{\texttt{MPC}}( {\bf x}_k, {\bf u}_k, {\bf r}_k  ) =
 || {\bf x}_k - {\bf r}_k ||_{Q_r}^2  + || {\bf u}_k ||_{Q_u}^2 
\end{equation}
with reference states ${\bf r}_k$, and $\|{\bf a}\|_Q^2 = {\bf a}^T Q {\bf a}$ the weighted squared $2$-norm.
The predictions over the prediction horizon $N$ are obtained from the system dynamics  represented by a controllable linear state space model~\eqref{eq:mpc_example:x}.  The  system is also subject to  state~\eqref{eq:mpc_example:xb} and input~\eqref{eq:mpc_example:ub} constraints, where $g({\bf x}_k): \mathbb{R}^{n_x} \to \mathbb{R}^{n_g} $ and $h({\bf u}_k): \mathbb{R}^{n_u} \to \mathbb{R}^{n_h} $ are differentiable functions. 
% In their simplest form are given by linear inequalities $ {\bf M} {\bf x}_k \le {\bf m}$, and $ {\bf L} {\bf x}_k \le {\bf l}$, respectively.

\begin{assumption}
\label{assume:ctrl}
The linear system dynamics model ${\bf x}_{k+1} = {\bf A}{\bf x}_{k}   + {\bf B }{\bf u}_{k}  $ is controllable.
\end{assumption}

Typically, the problem~\eqref{eq:mpc_example} is solved online via constrained optimization solvers, which is often referred to as implicit MPC. 
This approach is based on receding horizon control (RHC) principle, where new system measurements are obtained at each sampling instant and used as input parameters for the constrained optimization solver that computes optimal control trajectory ${\bf U} = {{\bf u}_0, \ldots, {\bf u}_{N-1}}$.
The main advantage of the implicit MPC are  constraints handling capabilities and stability guarantees. However, the need to
compute solutions to the constrained optimization problem 
in real-time  prohibits the application of implicit MPC for systems with fast sampling rates, or with limited computational resources.

\subsection{Explicit Model Predictive Control} \label{sec:empc}

% two fundamental approaches: sensitivity analysis and parametric programming \\
The real-time implementation of MPC typically requires an online solution of the corresponding constrained optimization problem~\eqref{eq:mpc_example}. This approach corresponds to the so-called implicit MPC, whose main limiting factor is higher online computational requirements that might be prohibitive for many real-world applications with limited computing resources.
Explicit MPC~\cite{TonEtal:aut:01,BemEtal:aut:02,borrelli:2017:book} represents an alternative approach based on an offline solution of the corresponding multi-parametric programming problem~\cite{MPT3:2013,kvasnica:vdm} obtained by reformulation of the original MPC problem~\eqref{eq:mpc_example}.
A generic form of the multi-parametric programming problem is given as:
\begin{subequations}
\label{eq:param_prog}
    \begin{align}
 \min_{{\bf U}} \ & J({\bf U}, {\boldsymbol \xi}) & 
 \label{eq:param_prog:Q} \\ 
  \text{s.t.} \ &  c({\bf U}, {\boldsymbol \xi}) \leq {\bf 0},   \label{eq:param_prog:con} \\
  \ & {\boldsymbol \xi} \in \Xi \subset \mathbb{R}^n
\end{align}
\end{subequations}
Where ${\bf U}$ represents a vector of optimized variables, and ${\boldsymbol\xi}$ is a vector of parameters, corresponding to a set of admissible state measurements, constraint bounds, and reference signals in the original problem~\eqref{eq:mpc_example}.
The offline solution of the parametric problem~\eqref{eq:param_prog}
yields an explicit state feedback control policy  $ {\bf U} = \pi(\boldsymbol \xi)$
which for linear constraints $c({\bf U}, {\boldsymbol \xi})$ with quadratic objective $ J({\bf U}, {\boldsymbol \xi})$ has the piecewise affine form:
\begin{equation}
    \label{eq:pwa_law}
    \pi(\boldsymbol \xi) = \alpha_r \boldsymbol \xi + \beta_r, \ \ \text{if} \ \boldsymbol \xi \in \mathcal{R}_r
\end{equation}
where $\alpha_r \in \mathbb{R}^{n_U \times n_{\xi}}$ and $\beta_r \in \mathbb{R}^{n_U}$ specify local affine laws
defined over polytopic regions $\mathcal{R}_r$ partitioning the parametric space $\boldsymbol \Xi$. 
Thus the control policy~\eqref{eq:pwa_law} represents 
a pre-computed solution of the original MPC problem~\eqref{eq:mpc_example}.

Hence the main advantage of the explicit MPC relative to implicit MPC lies in the fast, optimization-free online evaluation of the PWA control policy~\eqref{eq:pwa_law}.
Unfortunately, existing solutions~\cite{TonEtal:aut:01,gupta2011novel,MPT3:2013} of the parametric programming problem~\eqref{eq:param_prog}, as mentioned in Section \ref{sec:eMPC1}, scale geometrically with the number of constraints. Thus effectively limiting the applicability of explicit MPC only to small-scale problems with only a small number of decision variables over short prediction horizons.

\section{Differentiable Predictive Control}

\begin{figure}[htb!]
    \centering
     \includegraphics[width=.5\textwidth]{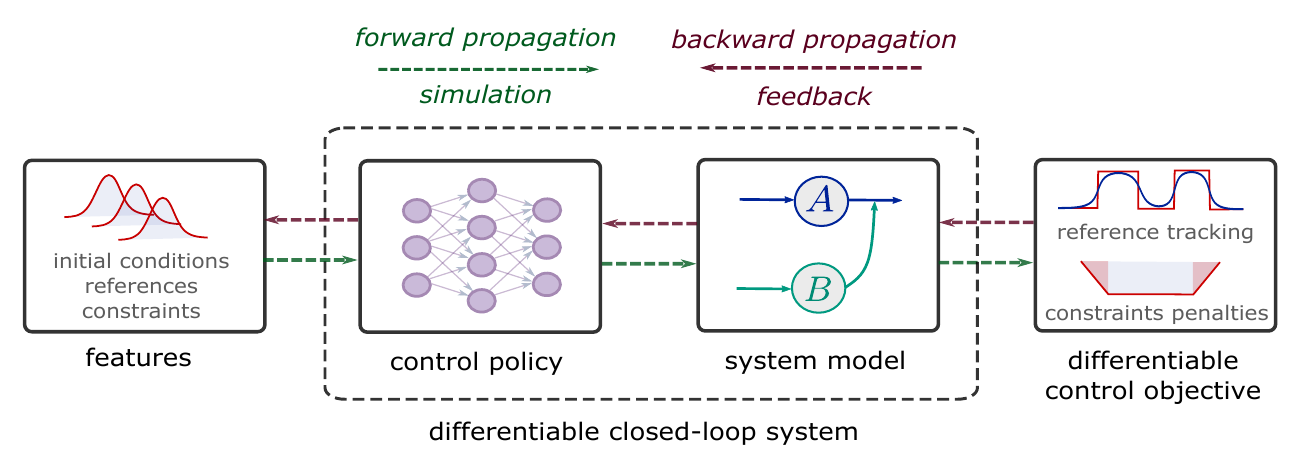}
    \caption{Conceptual computational graph of the proposed differentiable predictive control (DPC) method.}
    \label{fig:DPC_concept}
\end{figure}

We present a new perspective on the solution of parametric optimal control problems by leveraging the connection between parametric programming and data-driven differentiable programming.  
In particular, we introduce a differentiable predictive control (DPC) method based on representing the constrained optimal control problem as a differentiable program and using automatic differentiation to obtain direct policy gradients for gradient-based optimization.
The  proposed model-based DPC methodology is illustrated in Fig.~\ref{fig:DPC_concept}.
As a first step, the system's state-space model is obtained from system identification or physics-based modeling.
Next, we construct a differentiable closed-loop system model by combining the state space model and neural control policy in a computational graph. 
Finally, the neural control policy is optimized via backpropagation of the gradients of the MPC objective function and constraints penalties, formulated as soft constraints, through the unrolled closed-loop system dynamics. 

\subsection{Differentiable Predictive Control Problem Formulation}
\label{sec:dpc}

We consider the following parametric optimal control problem:
\begin{subequations}
\label{eq:DPC}
    \begin{align}
 \min_{{\bf W}} & \frac{1}{mN} \sum_{i=1}^{m} \sum_{k=0}^{N-1}  \big( \ell_{\texttt{MPC}}( {\bf x}_k^i, {\bf u}_k^i, {\bf r}_k^i ) +  &  \label{eq:DPC:objective1}\\
 & p_x(h({\bf x}_k^i, {\bf p_{h}}_k^i))  +   p_u(g({\bf u}_k^i, {\bf p_{g}}_k^i))  \big) & 
 \label{eq:DPC:objective} \\ 
  \text{s.t.} \ &   {\bf x}_{k+1}^i = {\bf A} {\bf x}_k^i + {\bf B } {\bf u}_k^i, \  k \in \mathbb{N}_{0}^{N-1}   \label{eq:dpc:x}  & \\
 \  & {\bf u}_k^i = \pi_{ {\bf W}}({\bf x}_k^i, \boldsymbol \xi_k^i)  \label{eq:dpc:pi} \\
      \ &  {\bf x}_0^i \in \mathbb{X} \subset \mathbb{R}^{n_{x}} \label{eq:dpc:x0} \\ 
  \ &\boldsymbol \xi_k^i = \{{\bf r}_k^i, {\bf p_{h}}_k^i, {\bf p_{g}}_k^i\} \in \Xi \subset \mathbb{R}^{n_{\xi}} \label{eq:dpc:xi}
\end{align}
\end{subequations}
The  DPC loss function is  
composed of the parametric MPC objective $ \ell_{\texttt{MPC}}( {\bf x}_k, {\bf u}_k, {\bf r}_k): \mathbb{R}^{n_x + n_u + n_{r}} \to \mathbb{R}  $,
and penalties of parametric constraints  $p_x(h({\bf x}_k, {\bf p_{h}}_k): \mathbb{R}^{n_h + n_{p_{h}}} \to \mathbb{R}   $, and
$ p_u(g({\bf u}_k, {\bf p_{g}}_k)): \mathbb{R}^{n_g + n_{p_{g}}} \to \mathbb{R} $.
The formulation~\eqref{eq:DPC} implemented as a differentiable program allows us to obtain a data-driven solution of the corresponding parametric programming problem~\eqref{eq:param_prog}
by differentiating the loss function~\eqref{eq:DPC:objective1} and \eqref{eq:DPC:objective} backwards through the parametrized closed-loop dynamics given by the system model~\eqref{eq:dpc:x}  and neural control policy~\eqref{eq:dpc:pi}.
This formulation allows one to use the stochastic gradient descent and its variants to minimize the loss function~\eqref{eq:DPC:objective1} and \eqref{eq:DPC:objective}
over a distribution of control parameters~\eqref{eq:dpc:xi} and initial conditions~\eqref{eq:dpc:x0} sampled from the synthetically generated
training dataset $\Xi$, where $m$ represents the total number of parametric scenario samples, and $i$ denotes the index of the sample.
In the following paragraphs we elaborate on key components of the proposed DPC problem formulation~\eqref{eq:DPC}.

\subsubsection{Neural control policy}
One major difference between implicit MPC~\eqref{eq:mpc_example} and DPC~\eqref{eq:DPC} is that the MPC solution
returns an optimal sequence of the control actions ${\bf U} = {{\bf u}_0, \ldots, {\bf u}_{N-1}}$,
while solving DPC yields optimized weights and biases ${\bf W}= \{\mathbf{H}_1, \ldots \mathbf{H}_{L}, 
\mathbf{b}_1, \ldots \mathbf{b}_{L}\}$ parametrizing the explicit neural control policy $\pi_{ {\bf W}}: \mathbb{R}^{n_x + n_{\xi}} \rightarrow \mathbb{R}^{n_u}$ given as:
\begin{subequations}
    \label{eq:dnn}
    \begin{align}
   \boldsymbol{\pi}_{ {\bf W}}({\bf x}_k, \boldsymbol {\xi}_k) & =  \mathbf{H}_{L}  \mathbf{z}_L + \mathbf{b}_{L} \\
    \mathbf{z}_{l} &= \boldsymbol\sigma(\mathbf{H}_{l-1} \mathbf{z}_{l-1} + \mathbf{b}_{l-1})  \label{eq:dnn:layer}\\
    \mathbf{z}_0 &= [ {\bf x}_k, \boldsymbol {\xi}_k ]
 \end{align}
\end{subequations}
Where $\mathbf{z}_i$ are hidden states, $\mathbf{H}_i$, and $\mathbf{b}_i$
represent weights and biases of the $i$-th layer, respectively.
The  activation layer $\boldsymbol\sigma: \mathbb{R}^{n_z} \rightarrow \mathbb{R}^{n_z}$ is given as the element-wise application of a differentiable univariate function $\sigma: \mathbb{R} \rightarrow \mathbb{R}$.

In the nominal case, we define a full state feedback policy as ${\bf z}_0 = {\bf x}_k$.
In the extended case, we can consider policy parametrized by
  a vector of state measurements ${\bf x}_k$, and parameter vector $\boldsymbol \xi_k = [ {\bf r}_k, {\bf p_{h}}_k, {\bf p_{g}}_k ]$ consisting of previews of the reference signals, and  state and input constraints parameters, respectively.
Thus allowing for generalization across tasks with full reference preview, and dynamic constraints handling capabilities as in the case of classical implicit MPC.

\subsubsection{Differentiable objectives and constraints}
The DPC problem~\eqref{eq:DPC}  is  solved by differentiating the parametric MPC loss in the Lagrangian form. Here, the first term of the objective gives a main performance metric, e.g., reference tracking loss
$\ell_{\texttt{MPC}}( {\bf x}_k, {\bf u}_k, {\bf r}_k ) =
 || {\bf x}_k - {\bf r}_k ||_{Q_r}^2  + || {\bf u}_k ||_{Q_u}^2 $ parametrized by ${\bf r}_k$.
While the last two terms correspond to penalty functions~\cite{BoyVan:ConOpt:04,BenNem:opt:01} imposed on state  and control action constraints parametrized by ${\bf p_{h}}_k$, and ${\bf p_{g}}_k$,  respectively. 
Specifically, the parametric constraints penalties used in this paper are given as:
\begin{subequations}
\label{eq:ReLU_ineq}
    \begin{align}
    p_x(h({\bf x}_k, {\bf p_{h}}_k)) & =
   Q_h \frac{1}{n_h}  ||
    \texttt{ReLU}(h_j({\bf x}_k, {\bf p_{h}}_k))||_l \\
p_u(g({\bf u}_k, {\bf p_{g}}_k ))  & = Q_g \frac{1}{n_g}  || \texttt{ReLU}(g_j({\bf u}_k, {\bf p_{g}}_k))||_l
 \end{align}
 \end{subequations}
% \begin{subequations}
% \label{eq:ReLU_ineq}
%     \begin{align}
%     p_x(h({\bf x}_k)) & =
%     \frac{1}{n_h}  \sum_{j=1}^{n_h} 
%     \texttt{ReLU}(h_j({\bf x}_k)), \\
% p_u(g({\bf x}_k) )  & = \frac{1}{n_g}  \sum_{j=1}^{n_g}  \texttt{ReLU}(g_j({\bf x}_k)),
%  \end{align}
%  \end{subequations}
 with $\texttt{ReLU}$ standing for rectifier linear unit function,
$Q_x$ and $Q_u$ representing scalar penalty weights,
 $n_h$ and  $n_g$ defining the total number of state and input constraints, and $l$  representing the penalty norm.

 \begin{assumption}
\label{assume:loss_con_diff}
The parametric control objective function $\ell_{\texttt{MPC}}( {\bf x}_k, {\bf u}_k, {\bf r}_k)$, and state and input constraints penalties $ p_x(h({\bf x}_k, {\bf p_{h}}_k))$ and  $p_u(g({\bf u}_k, {\bf p_{g}}_k))$ are differentiable  at almost every point in their domain.
\end{assumption}

\subsubsection{Differentiable closed-loop system}

\begin{figure*}[htb]
    \centering
     \includegraphics[width=.99\textwidth]{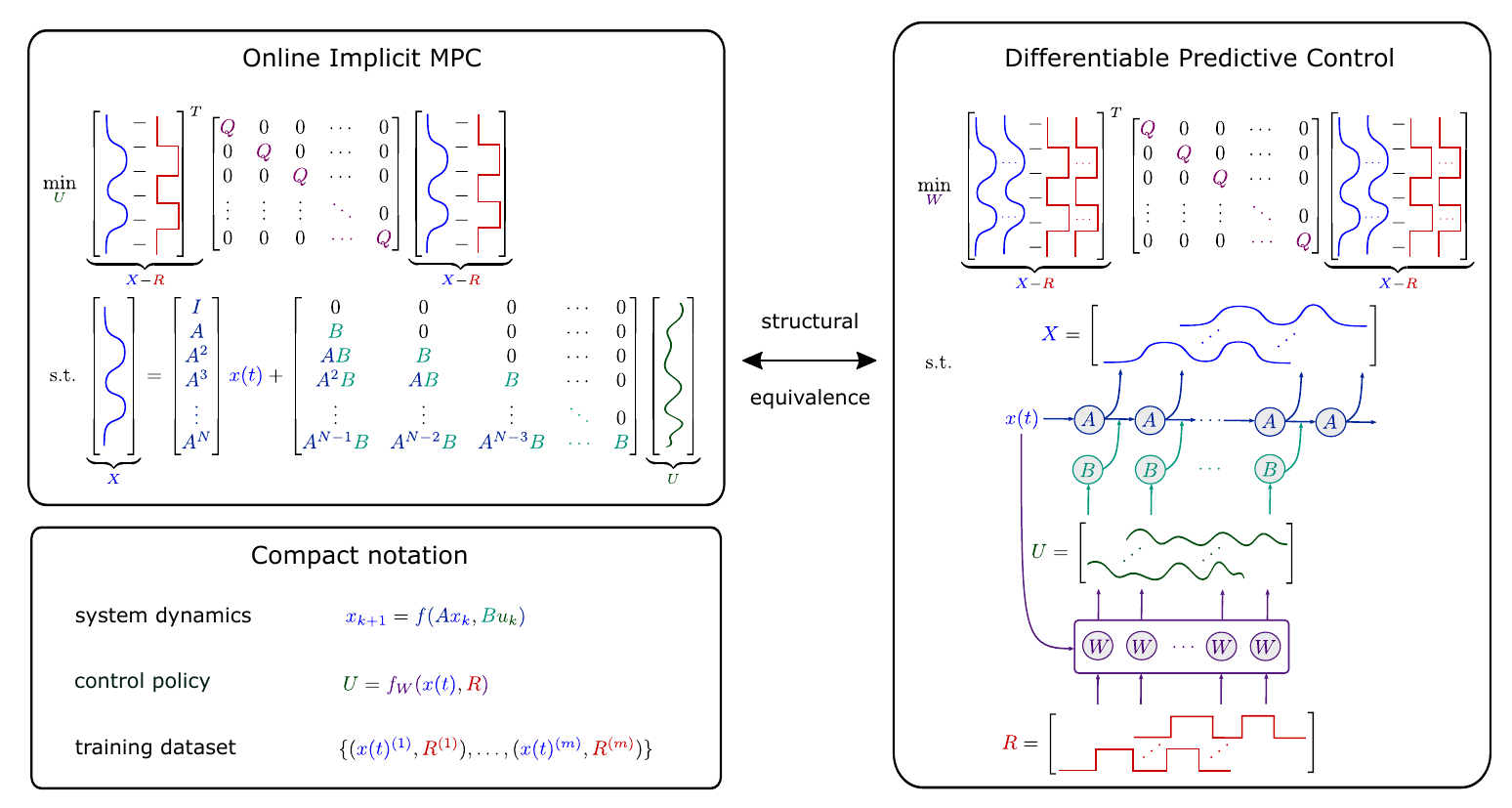}
    \caption{Structural equivalence of implicit MPC in a dense form (single shooting formulation) with the proposed differentiable predictive control (DPC). The unrolled computational graph of the DPC (right) is equivalent 
    with the single shooting formulation of the MPC system dynamics constraints in the standard matrix notation (left). 
    }
    \label{fig:MPC_structure}
\end{figure*}

The core idea of DPC is based on the parametrization of the differentiable closed-loop system composed of differentiable system dynamics model and neural control policy~\eqref{eq:dnn} as shown in Fig.~\ref{fig:DPC_concept}. In this paper, we assume that the dynamics is represented by a linear state space model.
In a nominal case with ${\bf z}_0 = {\bf x}_k$
we obtain a full state feedback formulation of the control problem~\eqref{eq:DPC}, where
 the system model~\eqref{eq:dpc:x} and control policy~\eqref{eq:dpc:pi} together 
define the closed-loop system dynamic as:
\begin{equation}
\label{eq:closed_loop}
    {\bf x}_{k+1} = {\bf A} {\bf x}_k + {\bf B } \pi_{ {\bf W}}({\bf x}_k) 
\end{equation}
To simulate $N$-step ahead closed-loop dynamics, we use the recurrent rollout of the system~\eqref{eq:closed_loop}. This is conceptually equivalent with a single shooting formulation of the MPC problem~\cite{Binder2001}. The resulting structural equivalence of the constraints of classical implicit MPC in a dense form with DPC is illustrated in Figure~\ref{fig:MPC_structure}.
Similarly to MPC, in the open-loop rollouts, the explicit DPC policy generates future control action trajectories over $N$-step prediction horizon given the feedback from the system dynamics model.
Then for the closed-loop deployment, we adopt the receding horizon control (RHC) strategy by applying only the first time step of the computed control action.

\subsubsection{Computing policy gradients of the parametric optimal control problem~\eqref{eq:DPC}}

A main advantage of having a differentiable closed-loop dynamics model, control objective function, and constraints is that it allows us to use
automatic differentiation (backpropagation through time~\cite{puskorius1994truncated}) to directly compute the policy gradient. 
In particular, by representing the problem~\eqref{eq:DPC} as a computational graph and  leveraging the chain rule, we can directly compute the gradients of the loss function $\mathcal{L}_{\texttt{DPC}}$ w.r.t. the policy parameters ${\bf W}$ as follows:
\begin{equation}
\label{eq:grad}
\begin{split}
    \nabla_{{\bf W}} \mathcal{L}_{\texttt{DPC}} =  \frac{ \partial\ell_{\texttt{MPC}}( {\bf x}, {\bf u}, {\bf r})}{\partial {\bf W}} + \\ \frac{ \partial p_x(h({\bf x}, {\bf p_{h}}))}{\partial {\bf W}} + 
    \frac{ \partial p_u(g({\bf u}, {\bf p_{g}}))}{\partial {\bf W}} = \\
    \frac{ \partial\ell_{\texttt{MPC}}( {\bf x}, {\bf u}, {\bf r})}{\partial {\bf x}}   \frac{ \partial {\bf x}}{\partial {\bf u}} \frac{ \partial {\bf u}}{\partial {\bf W}} +   \frac{ \partial\ell_{\texttt{MPC}}( {\bf x}, {\bf u}, {\bf r})}{\partial {\bf u}}  \frac{ \partial {\bf u}}{\partial {\bf W}} +  \\
    \frac{ \partial p_x(h({\bf x}, {\bf p_{h}}))}{\partial {\bf x}}  \frac{ \partial {\bf x}}{\partial {\bf u}} \frac{ \partial {\bf u}}{\partial {\bf W}} +   \frac{ \partial p_u(g({\bf u}, {\bf p_{g}}))}{\partial {\bf u}} \frac{ \partial {\bf u}}{\partial {\bf W}}
\end{split}
\end{equation}
Where $\frac{ \partial {\bf u}}{\partial {\bf W}}$ represent partial derivatives of the neural policy outputs w.r.t. its weights that are typically being computed in deep learning applications via backpropagation. The advantage of having gradients~\eqref{eq:grad} is that it allows us to use scalable stochastic gradient optimization algorithms such as AdamW~\cite{loshchilov2017decoupled} to solve the corresponding parametric optimal control problem~\eqref{eq:DPC} by direct offline optimization of the neural control policy.
In practice, we can compute the gradient of the DPC problem by using  automatic differentiation frameworks such as Pytorch~\cite{paszke2019pytorch}.

\begin{remark}
The advantage of a differentiable control problem~\eqref{eq:DPC} is that we can easily employ automatic differentiation to compute not only the policy gradients $ \nabla_{{\bf W}} \mathcal{L}_{\texttt{DPC}}$ but also the problem sensitivities to the system dynamics model parameters
$ \nabla_{{\bf A}} \mathcal{L}_{\texttt{DPC}}$, and $ \nabla_{{\bf B}} \mathcal{L}_{\texttt{DPC}}$, respectively. We will leverage this later in this section to obtain adaptive control updates via simultaneous system identification and policy optimization.
\end{remark}

\subsection{Differentiable Predictive Control Policy Optimization}
\label{sec:DPC_optim}
The proposed DPC method represents an offline learning method for obtaining explicit solutions of the underlying parametric optimal control problem~\eqref{eq:DPC}. The idea is based on a sampling of the parametric space to simulate the closed-loop dynamical system~\eqref{eq:closed_loop} in the forward pass and
using automatic differentiation in the reverse mode (backpropagation) to obtain the  gradients of the MPC objective and constraints~\eqref{eq:grad}
that are used for direct policy optimization via gradient-based optimization.
In the following subsections, we elaborate the aspects of the dataset sampling, forward pass architecture, loss function, and gradient-based policy optimization.

\subsubsection{DPC dataset}
First, we sample the initial conditions
of the problem ${\bf x}_0$ to generate a  set $\mathbb{X}$.
Next we populate the dataset of control parameters $\Xi = \{ \boldsymbol \xi_1, \ldots, \boldsymbol \xi_N\}$
by sampling  reference, and constraints parameters
over an $N$-step horizon window with a given distribution, e.g., randomly uniform. In particular, at each time step we consider the following  parameter set $\boldsymbol \xi_k = [ {\bf r}_k, {\bf p_{h}}_k, {\bf p_{g}}_k]$. 
Alternatively, the distribution of the training data can be obtained from measured trajectories of the controlled system or specified by a domain expert.

\subsubsection{Forward pass DPC}
The forward pass of the DPC computational graph is defined in the   Algorithm~\ref{algo:MPCpolicy}.
Here, the neural control policy~\eqref{eq:dnn} is evaluated on line~\ref{algo:MPCpolicy:pol}. 
In line~\ref{algo:MPCpolicy:x_plus}, the system model is integrated forward in time to simulate the evolution of the states. 
The violations of the state and control action constraints are computed on line~\ref{algo:MPCpolicy:x_s}, and~\ref{algo:MPCpolicy:u_s}, respectively. 
The control loss is calculated on line~\eqref{algo:MPCpolicy:loss}.
 And finally, the forward pass of the policy returns 
 computed control actions, predicted states, and slacks representing constraints violations, and control loss values. 
\begin{algorithm}
  \caption{Forward pass of a single time step  evaluation of DPC computational graph.}\label{algo:MPCpolicy}
  \begin{algorithmic}[1]
    \Function{DPC}{${\bf x}_k, \boldsymbol \xi_k$}
    \State  $\boldsymbol \xi_k = [ {\bf r}_k, {\bf p_{h}}_k, {\bf p_{g}}_k]$          \Comment{problem parameters} \label{algo:MPCpolicy:xi}
    \State  ${\bf u}_k =  \pi_{{\bf W}}({\bf x}_k,{\boldsymbol \xi_k}) $   \Comment{neural policy evaluation}  \label{algo:MPCpolicy:pol}
 \label{algo:MPCpolicy:u_ub}
    \State ${\bf {x}}_{k+1} = {\bf A} {\bf x}_{k} +  {\bf B} {\bf u}_{k}$
    \Comment{state update} \label{algo:MPCpolicy:x_plus}
    \State  ${\bf s}_k^h = p_x(h({\bf x}_k, {\bf p_{h}}_k)) $     \Comment{state constraints penalties}
    \label{algo:MPCpolicy:x_s}
        \State  ${\bf s}_k^g = p_u(g({\bf u}_k, {\bf p_{g}}_k)) $     \Comment{input constraints  penalties}
    \label{algo:MPCpolicy:u_s}
        \State  ${\bf s}_k^l = \ell_{\texttt{MPC}}( {\bf x}_k, {\bf u}_k, {\bf r}_k ) $     \Comment{control objective}
    \label{algo:MPCpolicy:loss}
    \State\Return ${\bf u}_k,  {\bf {x}}_{k+1}, {\bf s}_k^h, {\bf s}_k^g, {\bf s}_k^l$ \label{algo:MPCpolicy:return}
    \EndFunction
  \end{algorithmic}
\end{algorithm}

\subsubsection{DPC loss function}
The trainable  parameters ${\bf W}$ of the \textsc{DPC} are   optimized w.r.t. the following weighted multi-objective loss function:
\begin{equation}
\label{eq:policy_loss}
\begin{split}
\mathcal{L}_{\texttt{DPC}}(\mathbb{X},\Xi| \text{DPC}({\bf x}_k, \boldsymbol \xi_k)) =  \\ \frac{1}{mN} \sum_{i=1}^{m}  \sum_{k=0}^{N-1}   \big( ||{{\bf s}_{k}^l}^i||^2_{Q_l} + 
  ||{{\bf s}_{k}^h}^i||^2_{Q_h}  + ||{{\bf s}_{k}^g}^i||^2_{Q_g} \big)
\end{split}
\end{equation}
Where, the first term evaluates the control performance loss, while the last two terms evaluate  state and input constraints violations, respectively, over the 
distributions of initial conditions sampled from the set $\mathbb{X}$, and control parameters sampled from the set $\Xi$.
The relative importance of the individual terms can be tuned  by weight factors $Q_l$, $Q_h$, and $Q_g$.

\subsubsection{DPC policy learning algorithm}
Now combining the individual components, the DPC policy optimization algorithm is summarized in Algorithm~\ref{algo:DPC_optim}.
The system dynamics model is required to instantiate the  
 computational graph of the DPC problem as specified in Algorithm~\ref{algo:MPCpolicy}.
 The policy gradients $ \nabla_{{\bf W}} \mathcal{L}_{\texttt{DPC}} $ are obtained by differentiating the DPC loss function  $\mathcal{L}_{\texttt{DPC}}$ over the distribution of initial state conditions and problem parameters
 sampled from the given training datasets $\mathbb{X}$ and    $\Xi $, respectively.
 The computed policy gradients now allow us
 to perform direct policy optimization via a gradient-based optimizer $\mathbb{O}$.
Thus the presented procedure introduces a generic approach for   data-driven solution of model-based
parametric optimal control problem~\eqref{eq:DPC}
with constrained neural control policies.
\begin{algorithm}[ht!]
  \caption{DPC policy optimization.}\label{algo:DPC_optim}
  \begin{algorithmic}[1]
  \State \textbf{input} training datasets of sampled initial conditions $\mathbb{X}$ and problem parameters  $\Xi $
  \State \textbf{input} system dynamics model ${\bf x}_{k+1} = {\bf A} {\bf x}_k + {\bf B } {\bf u}_k$
 \State \textbf{input}  DPC policy forward pass as given in Algorithm~\ref{algo:MPCpolicy}
   \State \textbf{input}  DPC loss  $\mathcal{L}_{\texttt{DPC}}$~\eqref{eq:policy_loss}
 \State \textbf{input} optimizer $\mathbb{O}$
 \State \textbf{differentiate}  DPC loss $\mathcal{L}_{\texttt{DPC}}$~\eqref{eq:policy_loss} to obtain the policy gradient $ \nabla_{{\bf W}} \mathcal{L}_{\texttt{DPC}} $~\eqref{eq:grad}
 \State \textbf{learn} policy $\pi_{ {\bf W}}$  via optimizer  $\mathbb{O}$ using gradient $ \nabla_{{\bf W}} \mathcal{L}_{\texttt{DPC}} $
\State \textbf{return} optimized policy $\pi_{ {\bf W}}$
  \end{algorithmic}
\end{algorithm}

\begin{remark}
From a reinforcement learning (RL) perspective, 
the DPC loss  $\mathcal{L}_{\texttt{DPC}}$ can be seen as a reward function, with $ \nabla_{{\bf W}} \mathcal{L}_{\texttt{DPC}} $ representing a deterministic policy gradient.  
The main difference compared with actor-critic RL algorithms is that 
in DPC the reward function is fully parametrized by a closed-loop system dynamics model, control objective, and constraints penalties. The model-based approach avoids approximation errors in reward functions making DPC more sample efficient than model-free RL  algorithms.
\end{remark}

\subsection{Adaptive DPC Policy Optimization}
\label{sec:DPC_optim_adaptive}
An adaptive DPC approach enables online updates of the constrained control policy and system dynamics model parameters. We can straightforwardly achieve this adaptive control capability by the following modifications to the DPC method.
Training of both the control policy and the dynamics model now requires initial state conditions $\mathbb{X}$ and  control parameters  $\Xi$ datasets alongside the system identification dataset, $\mathbb{D}^{\text{ID}}$.
\begin{assumption}
For training the unknown linear system dynamics model ${\bf {x}}_{k+1} = {\bf \tilde{A}} {\bf x}_{k} +  {\bf \tilde{B}} {\bf u}_{k} $, 
we assume access to a representative dataset of the controlled system dynamics in the form of input-output time series $\mathbb{D}^{\text{ID}} = \{ {\mathbf{U}^{\text{ID}}}, \mathbf{X}^{\text{ID}} \}$.
\end{assumption}

For learning the system dynamics model matrices ${\bf \tilde{A}}$ and ${\bf \tilde{B}}$ we use an additional loss term penalizing the residuals between predicted states ${\bf \hat{x}}_k^{\text{ID}}$ and measured states ${\bf x}_k^{\text{ID}}$: 
\begin{equation}
\label{eq:sys_id_loss}
    \ell_{\texttt{ID}}( {\bf x}_k^{\text{ID}}, {\bf \hat{x}}_k^{\text{ID}} ) = ||{\bf x}_k^{\text{ID}} - {\bf \hat{x}}_k^{\text{ID}} ||^2_2
\end{equation}

\subsubsection{Forward pass adaptive DPC architecture}
 The modified  forward pass architecture is defined by Algorithm~\ref{algo:model_MPCpolicy}, where
 the control policy (line \ref{algo:model_MPCpolicy:pol})
 is learned together  with system model dynamics (line \ref{algo:model_MPCpolicy:x_plus_ID}).
 Here, the examples of state trajectories $\mathbf{X}^{\text{ID}} = \{{\bf x}_0^{\text{ID}}, \ldots {\bf x}_N^{\text{ID}} \}$  are used as targets for system identification loss evaluated on line~\ref{algo:model_MPCpolicy:id_loss}. 
 While 
 randomly sampled states ${\bf x}_0$  are used to initialize the rollouts of the second instance of the system model (line~\ref{algo:model_MPCpolicy:x_plus}) to guide the control policy update through control loss, and constraints penalties that are evaluated as in Algorithm~\ref{algo:MPCpolicy}.
 Analogously, ${\bf u}_k^{\text{ID}}$, and ${\bf x}$ are control inputs associated with the system identification and policy learning, respectively. Where the former one is obtained from the measured trajectories stored in the system identification training set $\mathbb{D}^{\text{ID}} $ and the latter one is generated by the neural control policy.
\begin{algorithm}
 \caption{Forward pass of a single time step evaluation of  DPC computational graph with adaptive system model updates.}\label{algo:model_MPCpolicy}
 \begin{algorithmic}[1]
    \Function{aDPC}{${\bf x}_k,  {\bf x}_k^{\text{ID}}, {{\bf u}}_k^{\text{ID}}, \boldsymbol \xi_k$}
        \State  $\boldsymbol \xi_k = [ {\bf r}_k, {\bf p_{h}}_k, {\bf p_{g}}_k]$          \Comment{problem parameters} \label{algo:model_MPCpolicy:xi}
    \State  ${\bf u}_k =  \pi_{{\bf W}}({\bf x}_k, {\boldsymbol \xi_k}) $   \Comment{control policy evaluation}  \label{algo:model_MPCpolicy:pol}
    \label{algo:model_MPCpolicy:u_s1}
    \State ${\bf {x}}_{k+1} = {\bf \tilde{A}} {\bf x}_{k} +  {\bf \tilde{B}} {\bf u}_{k} $
    \Comment{control state update} \label{algo:model_MPCpolicy:x_plus}
    \State ${\bf \hat{x}}_{k+1}^{\text{ID}} =  {\bf \tilde{A}} {\bf x}_{k}^{\text{ID}} +  {\bf \tilde{B}} {\bf u}_{k}^{\text{ID}} $
    \Comment{system ID state update} \label{algo:model_MPCpolicy:x_plus_ID}
 \State  ${\bf s}_k^h = p_x(h({\bf x}_k, {\bf p_{h}}_k)) $     \Comment{state constraints penalties}
    \label{algo:model_MPCpolicy:x_s}
        \State  ${\bf s}_k^g = p_u(g({\bf u}_k, {\bf p_{g}}_k)) $     \Comment{input constraints  penalties}
    \label{algo:model_MPCpolicy:u_s2}
        \State  ${\bf s}_k^l = \ell_{\texttt{MPC}}( {\bf x}_k, {\bf u}_k, {\bf r}_k ) $     \Comment{control objective}
    \label{algo:model_MPCpolicy:loss}
  \State  ${\bf s}_k^{d} = \ell_{\texttt{ID}}( {\bf x}_k^{\text{ID}}, {\bf \hat{x}}_k^{\text{ID}} ) $     \Comment{system ID objective}
    \label{algo:model_MPCpolicy:id_loss}
    \State\Return ${\bf u}_k,  {\bf \tilde{x}}_{k+1},  {\bf \hat{x}}_{k+1}^{\text{ID}}, {\bf s}_k^h, {\bf s}_k^g, {\bf s}_k^l, {\bf s}_k^{d} $ \label{algo:model_MPCpolicy:return}
    \EndFunction
 \end{algorithmic}
\end{algorithm}

\subsubsection{Adaptive DPC loss function}
To allow for model updates, the DPC policy learning loss function~\eqref{eq:policy_loss}
is extended with a system identification term with weight $Q_{d}$ that penalizes the model deviations from measured states as follows:
\begin{equation}
\label{eq:SysID_policy_loss}
\begin{split}
\mathcal{L}_{\texttt{aDPC}}(\mathbb{X}, \mathbb{D}^{\text{ID}}, \Xi| \text{aDPC}({\bf x}_k,  {\bf x}_k^{\text{ID}}, {{\bf u}}_k^{\text{ID}}, \boldsymbol \xi_k)) =  \\ \frac{1}{mN} \sum_{i=1}^{m}  \sum_{k=0}^{N-1}   \big( ||{{\bf s}_{k}^l}^i||^2_{Q_l} + 
  ||{{\bf s}_{k}^h}^i||^2_{Q_h}  + ||{{\bf s}_{k}^g}^i||^2_{Q_g} +
   || {{\bf s}_k^{d}}^i ||^2_{Q_{d}} \big)
\end{split}
\end{equation}

\subsubsection{Adaptive DPC policy learning algorithm}
To jointly optimize the unknown model and policy parameters  ${\bf \tilde{A}}, {\bf \tilde{B}}, {\bf W}$, the gradient updates are performed using the aggregate loss~\eqref{eq:SysID_policy_loss} associated with the parallel system identification and control policy trajectories.
The modified adaptive DPC policy optimization algorithm is then given in Algorithm~\eqref{algo:aDPC_optim}.
\begin{algorithm}
  \caption{Adaptive DPC policy optimization.}\label{algo:aDPC_optim}
  \begin{algorithmic}[1]
  \State \textbf{input} training datasets of initial conditions $\mathbb{X}$ 
  \State \textbf{input} training datasets of control parameters $\Xi$ 
    \State \textbf{input} training datasets for system identification  $\mathbb{D}^{\text{ID}}$
 \State \textbf{input} adaptive DPC forward pass as given in Algorithm~\ref{algo:model_MPCpolicy}
  \State \textbf{input} adaptive DPC loss  $\mathcal{L}_{\texttt{aDPC}}$~\eqref{eq:SysID_policy_loss}
 \State \textbf{input} optimizer $\mathbb{O}$
  \State \textbf{differentiate}  adaptive DPC loss $\mathcal{L}_{\texttt{aDPC}}$~\eqref{eq:SysID_policy_loss} to obtain the gradients $ \nabla_{{\bf \tilde{A}}, {\bf \tilde{B}}, {\bf W}} \mathcal{L}_{\texttt{aDPC}} $
 \State \textbf{learn} system dynamics model ${\bf {x}}_{k+1} = {\bf \tilde{A}} {\bf x}_{k} +  {\bf \tilde{B}} {\bf u}_{k} $ and policy $\pi_{ {\bf W}}$  via optimizer  $\mathbb{O}$ using gradients $ \nabla_{{\bf \tilde{A}}, {\bf \tilde{B}}, {\bf W}} \mathcal{L}_{\texttt{aDPC}} $
\State \textbf{return} learned model matrices ${\bf \tilde{A}}$, ${\bf \tilde{B}}$ and  policy $\pi_{ {\bf W}}$
  \end{algorithmic}
\end{algorithm}

\begin{remark}
To avoid catastrophic forgetting during online learning, we can update only a subset of the policy and model parameters while fixing the rest. This allows for adapting to new data while preserving most of the learned dynamics from the offline learning phase.
\end{remark}

% https://engineering.utsa.edu/ataha/wp-content/uploads/sites/38/2017/10/MPC_Intro.pdf

\section{Stability and Constraints Satisfaction of Differentiable Predictive Control}
\label{sec:stability_DPC}
Closed-loop system stability guarantees and constraints satisfaction are premier features of MPC. 
In this section, we elaborate on the stability of the DPC method by leveraging the structural equivalence with MPC, which allows us to  apply stability promoting methods developed for MPC in the context of DPC.
Moreover, we provide sufficient stability conditions of the closed-loop systems with linear state space models of the system dynamics and full state feedback neural control policies.
Finally, we provide a practical stability and constraints satisfaction verification method that can be used as a safety certificate for the neural policies trained using the DPC policy optimization Algorithm~\ref{algo:DPC_optim} or  Algorithm~\ref{algo:aDPC_optim}, respectively.

\subsection{Probabilistic Stability and Constraints Satisfaction}

In this section, we present a probabilistic verification method for closed-loop stability and constraints satisfaction guarantees of the proposed DPC policy  optimization Algorithm~\ref{algo:DPC_optim}.
In particular, we modify the probabilistic verification method developed by~\cite{Hertneck8371312} in the context of approximate MPC method.
We first consider the following assumptions.
\begin{assumption}
\label{assume:no_mismatch}
There is no model mismatch between prediction model~\eqref{eq:dpc:x} and controlled system dynamics.
\end{assumption}
\begin{assumption}
\label{assume:terminal_set}
There exists a local Lyapunov function $V_f( \mathbf{x}) = ||  \mathbf{x} ||_P^2$, a terminal set $\mathcal{X}_f := \{\mathbf{x} :
V_f( \mathbf{x}) \le \alpha_f \}$, and a  feedback control law $\boldsymbol{\pi}_{ {\bf W}}( {\bf x})$, such that  $\forall \mathbf{x} \in \mathcal{X}_f$ and given state and input constraints sets $\mathcal{X}$, and $\mathcal{U}$, the following holds:
\begin{subequations}
\begin{align}
    % V_f( {\bf A}\mathbf{x}  + {\bf B } {\bf K}_f(\mathbf{x}) )
    % \le  V_f(\mathbf{x}) - || \mathbf{x} ||_Q^2 - || {\bf K}_f(\mathbf{x}) ||_R^2 \\
        V_f( {\bf A}  + {\bf B } \boldsymbol{\pi}_{{\bf W}}(\mathbf{x})  )
    \le  V_f(\mathbf{x})  \\
    \mathbf{x} \in \mathcal{X}, \ \boldsymbol{\pi}_{{\bf W}}(\mathbf{x}) \in \mathcal{U}
\end{align}
\end{subequations}
\end{assumption}

\subsubsection{Performance indicator functions}
Inspired by~\cite{Hertneck8371312,Karg2021}, our probabilistic validation method is based on the closed-loop system~\eqref{eq:closed_loop} rollouts defined as:
\begin{equation}
    \mathbf{X}^i = \{ \mathbf{x}_0^i, \ldots,  \mathbf{x}_N^i :
    \mathbf{x}_0^i \in \mathcal{X}_{\text{feas}},    s.t.~\eqref{eq:closed_loop}\}
\end{equation}
Here $i$ is an index of the simulated trajectory, and $\mathcal{X}_{\text{feas}}$ represents a feasible set  containing states that satisfy the constraints of the problem~\eqref{eq:DPC}, i.e. $ \mathbf{x} \in \mathcal{X}, \  \pi_{ {\bf W}}({\bf x}) \in \mathcal{U}$.

For performance assessment, we define stability $I_s(\mathbf{X}^i)$ and constraints satisfaction $I_c(\mathbf{X}^i)$  indicator functions as:
\begin{subequations}
\label{eq:indicators}
\begin{align}
 I_{s}(\mathbf{X}^i) & = \begin{cases}
  1 ,&  \mathbf{x}_N^i \in  \mathcal{X}_f, \ \text{for} \ \mathbf{x}_N \in \mathbf{X}^i \\
    0,              & \text{otherwise}
\end{cases} \label{eq:Is} \\
 I_{c}(\mathbf{X}^i) & = \begin{cases}
  1 ,& \text{if } \mathbf{x} \in \mathcal{X}, \  \pi_{ {\bf W}}({\bf x})  \in \mathcal{U}, \ \forall \mathbf{x} \in \mathbf{X}^i \\
    0,              & \text{otherwise}
\end{cases} \label{eq:Ic}
\end{align}
\end{subequations}
The proposed indicator functions~\eqref{eq:indicators} 
indicate whether the learned control policy $\pi_{ {\bf W}}({\bf x}_k) $ generates state trajectories that satisfy given stability conditions~\eqref{eq:Is}, without violating state and input constraints~\eqref{eq:Ic}, respectively. 

\begin{remark}
The stability indicator function~\eqref{eq:Is} is a generic concept and as such can be modified to employ different stability indicators, such as  contraction constraint~\eqref{eq:state_contract_con}, stability condition~\eqref{eq:stability_condition}, evaluation of a learned global Lyapunov function~\cite{Donti_control2020}, or a set of closed-loop performance metrics as proposed in~\cite{Karg2021}.
\end{remark}

\begin{remark}
The advantage of the proposed DPC problem formulation~\eqref{eq:DPC}
in comparison with approximate MPC scheme~\cite{Hertneck8371312} is that it allows us to define direct indicator functions~\eqref{eq:indicators} for verifiable performance based on known constraints and closed-loop system dynamics~\eqref{eq:closed_loop}.
While in~\cite{Hertneck8371312} the indicator functions are based on approximation error of the control policy trained via imitation learning supervised by the original MPC controller without access to the closed-loop system dynamics or constraints.
\end{remark}

\subsubsection{Probabilistic guarantees}
We can now use the indicator functions~\eqref{eq:indicators}  to validate the set of $m$ trajectories $\mathbf{X}^i, \ i \in \mathbb{N}_1^m$, with sampled independent and identically distributed (iid) initial conditions $\mathbf{x}_0^i$ from a feasible set $\mathcal{X}_{\text{feas}}$.
To evaluate the given condition violations we define a weighted empirical risk:
\begin{equation}
\label{eq:weigh_empirical_risk}
  \tilde{\mu} =   \frac{1}{m} \sum_{i=1}^m ( \alpha I_{s}(\mathbf{X}^i)  + \beta I_{c}(\mathbf{X}^i) )
\end{equation}
where, coefficients $ \alpha +  \beta  = 1$ represent relative importance weights for stability and constraints satisfaction, respectively. By default we consider $ \alpha = 0.5$,  $ \beta = 0.5$.
We consider $\mu := \mathbb{P}[I_{s}(\mathbf{X}^i)=1 \And I_{c}(\mathbf{X}^i)=1] $  as a lower bound for the probability  of closed-loop stability and constraints satisfaction  of the given trajectory $\mathbf{X}^i$.

Similar to~\cite{Hertneck8371312}, we use the Hoeffding's inequality~\cite{Hoeffding1963} to estimate the risk $\mu $ from the empirical risk $\tilde{\mu}$. For further details see Lemma~\ref{lema:Hoeffding} adopted from~\cite{Hertneck8371312}.
 Hoeffding's inequality implies that 
the probability of the sampled trajectories  $\mathbf{X}^i$ satisfying stability conditions and all constraints is larger than the empirical risk $\tilde{\mu } $ minus the user defined risk tolerance $\epsilon$
with confidence $1-\delta $, or more compactly $\mathbb{P}[I(\mathbf{X}^i)=1] = \mu  \ge \tilde{\mu } -  \epsilon$.
Thus for a chosen confidence $\delta$ and risk lower bound $\mu_{\text{bound}}$ we can evaluate the empirical risk bound as given in~\cite{Hertneck8371312}:
\begin{equation}
\label{eq:mu_bound}
    \mu_{\text{bound}} \le \tilde{\mu} - \epsilon =
    \tilde{\mu} - \sqrt{-\frac{\ln{\frac{\delta}{2}}}{2m}}
\end{equation}
where $m$ represents number of samples.
In our context, the condition~\eqref{eq:mu_bound} allows us to define the following closed-loop performance validation Algorithm~\ref{algo:validate} that can be used in conjunction with DPC policy optimization Algorithm~\ref{algo:DPC_optim}, and Algorithm~\ref{algo:aDPC_optim}, respectively.
\begin{algorithm}[ht!]
  \caption{Closed-loop stability and constraints satisfaction validation of DPC policy optimization algorithm.}\label{algo:validate}
  \begin{algorithmic}[1]
  \State  \textbf{input} stability and constraints indicator functions~\eqref{eq:indicators}
   \State  \textbf{input} empirical risk weights $\alpha$, $\beta$
 \State  \textbf{input} confidence $\delta$ and risk lower bound $\mu_{\text{bound}}$ 
 \State  \textbf{input} number of initial samples $m$, maximum number of samples $ m_{\text{max}}$,  and integer increment $p$
 \State  \textbf{sample} $m$ iid initial conditions   $\mathbf{x}_0^i \in \mathcal{X}_{\text{feas}}, \ i \in \mathbb{N}_1^m$ \label{step:sample}
    \State  \textbf{train}  neural control policy via DPC policy optimization Algorithm~\ref{algo:DPC_optim} or Algorithm~\ref{algo:aDPC_optim} on sampled initial conditions from step~\ref{step:sample} \label{step:init}
  \State  \textbf{rollout}  closed-loop system~\eqref{eq:closed_loop} over $N$ steps for each initial condition $\mathbf{x}_0^i$ to obtain state trajectories $\mathbf{X}^i$
  \State \textbf{evaluate} indicator functions~\eqref{eq:indicators} for each $\mathbf{X}^i$
    \State \textbf{evaluate} weighted empirical risk~\eqref{eq:weigh_empirical_risk}  
    and  risk lower bound~\eqref{eq:mu_bound}
     \If{ risk lower bound~\eqref{eq:mu_bound} is not satisfied}
     \State  increased number of samples $m := m+p$  
    \If{$m \ge m_{\text{max}}$}
      \State validation failed, terminate procedure
     \Else 
        \State repeat from step~\ref{step:sample}
    \EndIf
  \Else
    \State validation passed, terminate procedure
  \EndIf
  \end{algorithmic}
\end{algorithm}
Finally, inspired by~\cite{Hertneck8371312},
we can formalize the stability guarantees for DPC policy optimization obtained by running the Algorithm~\ref{algo:validate} in the following Theorem~\ref{thm:validation}.
\begin{theorem}
\label{thm:validation}
Let Assumptions~\ref{assume:ctrl}, ~\ref{assume:no_mismatch}, and~\ref{assume:terminal_set} hold. Construct the differentiable predictive control problem~\eqref{eq:DPC} with given system dynamics model.
Initialize and run the Algorithm~\ref{algo:validate} with 
chosen hyperparameters $\alpha$,  $\beta$, $\delta$, $\mu_{\text{bound}}$, $m$, $m_{\text{max}}$, and $p$.
If the Algorithm~\ref{algo:validate} terminates with validation procedure passed, then with confidence of $1-\delta$ 
the trained control policy satisfies closed-loop stability
and constraint satisfaction for $\mu_{\text{bound}} \cdot 100\%$
sampled initial conditions.
\end{theorem}

\begin{proof}
If the Algorithm~\ref{algo:validate} terminates with validation procedure passed, then~\eqref{eq:mu_bound} must hold, then through Hoeffding's inequality
 it holds that $ \mathbb{P}[I_{s}(\mathbf{X}^i)=1 \And I_{c}(\mathbf{X}^i)=1] \ge \mu_{\text{bound}}$ is true with confidence of $1-\delta$. 
Or in other words, the stability and constraints are satisfied for  at minimum  $\mu_{\text{bound}} \cdot 100\%$ samples with $1-\delta$ confidence. \qed
\end{proof}

% \add{rewrite this section as discussion not as novelty}

% \add{formulate this subsection as difference with Allgover paper, avoid borrowing the structure of their argument}

% \add{restructure: start with subsecion C, define the difference with the other paper, 
% follow with section B supporting section C, and finish with section A as discussion/observations}

% \begin{remark}
% In this work, we assume iid uniformly distributed sampling of initial conditions. However, more sophisticated sampling approaches could be inspired by works~\cite{ReLU_linReg2014,Serra2017,hanin2019complexity}  focusing on the evaluation of the number of linear regions in deep neural networks.
% We leave this line of research for future work.
% \end{remark}

% \subsubsection{Probabilistic constraints satisfaction of DPC policies}
% \label{sec:con:verify}

% % use lemma in our paper to obtain probabilistic constraints satisfaction based on state space sampling
% % https://ieeexplore.ieee.org/stamp/stamp.jsp?tp=&arnumber=6760908
%  \begin{assumption}
% \label{assume:con_convex}
% State and input constraints  $ h({\bf x}_k)$ and  $ g({\bf u}_k)$ are convex functions.
% \end{assumption}

\subsection{Stability of Neural Closed-Loop Systems}
\label{sec:stability}

In this section, we derive 
 stability condition that can be used as a metric in the stability indicator function~\eqref{eq:indicators} to obtain the probabilistic guarantees as described in the previous section.
In particular,  we provide sufficient closed-loop stability conditions for linear state space models controlled by a full state feedback neural policies~\eqref{eq:closed_loop}. 
We start by showing that under weak assumptions on the neural policy architecture, the closed loop system~\eqref{eq:closed_loop} is equivalent to piecewise affine system.
Then we derive sufficient conditions for closed-loop stability based on contractions of PWA maps and Banach fixed point theorem.

% \subsection{Point-wise Linearization of the Closed-loop Dynamics}
\subsubsection{Neural closed-loop dynamics as piecewise affine systems}
\label{sec:LPV}

First, we make use of Lemma~\ref{lem:lpv} that shows the equivalence of deep feedforward neural networks with pointwise affine maps.
\begin{lemma}~\cite{drgona2021stochastic}
 \label{lem:lpv}
A deep feedforward neural network $\boldsymbol{\pi}_{ {\bf W}}( {\bf x})$~\eqref{eq:dnn} with activation
function $ \boldsymbol\sigma$ is equivalent to a pointwise affine map parametrized by $\mathbf{x}$:
\begin{align}
 \label{eq:lpv_dnn}
\boldsymbol{\pi}_{ {\bf W}}( {\bf x}) = \mathbf{H}_{\boldsymbol{\pi}}(\mathbf{x}) \mathbf{x} + \mathbf{b}_{\boldsymbol{\pi}}(\mathbf{x}).
\end{align}
Where  $\mathbf{H}_{\boldsymbol{\pi}}(\mathbf{x})$ is a parameter varying matrix constructed as:
\begin{equation}
  \label{eq:dnn_LPV}
   \mathbf{H}_{\boldsymbol{\pi}}(\mathbf{x})= \mathbf{H}_L \boldsymbol\Lambda_{\mathbf{z}_{L-1}} \mathbf{H}_{L-1} \ldots \boldsymbol \Lambda_{\mathbf{z}_{0}} \mathbf{H}_0 
 \end{equation}
And $\mathbf{b}_{\boldsymbol{\pi}}(\mathbf{x})$ is a parameter varying vector defined as:
  \begin{subequations}
  \label{eq:dnn_bias_recurence}
  \begin{align}
 \mathbf{b}_{\boldsymbol{\pi}}(\mathbf{x}) =  \mathbf{b}^{\boldsymbol{\pi}}_{L}  \\
  \mathbf{b}^{\boldsymbol{\pi}}_{l} := \mathbf{A}^{\boldsymbol{\pi}}_l \boldsymbol\Lambda^{\boldsymbol{\pi}}_{\mathbf{z}_{l-1}} \mathbf{b}^{\boldsymbol{\pi}}_{l-1}  +\mathbf{A}^{\boldsymbol{\pi}}_l\boldsymbol\sigma_{l-1}(\mathbf{0}) +\mathbf{b}_{l}, \ \ l \in \mathbb{N}_1^{L}
  \end{align}
    \end{subequations}
 where   $ \mathbf{b}^{\boldsymbol{\pi}}_{0} = \mathbf{b}_{0}$, and $l$ standing for layer index.
 The parameter varying diagonal matrix $\boldsymbol \Lambda_{\mathbf{z}_{l}}$ represents activation patterns generated by functions $ \boldsymbol\sigma$  at each layer and is given as:
   \begin{equation}
\label{eq:lambda_matrix}
\begin{split}
    \boldsymbol\sigma(\mathbf{z})  =  \begin{bmatrix}
   \frac{\sigma(z_1) -\sigma(0)}{z_1} &  & \\
    & \ddots & \\
     &  &  \frac{\sigma(z_n)-\sigma(0)}{z_n} 
  \end{bmatrix}\mathbf{z} + \begin{bmatrix}\sigma(0)\\\vdots \\\sigma(0)\end{bmatrix} = \\ \boldsymbol \Lambda_{\mathbf{z}}  \mathbf{z} + \boldsymbol\sigma(\mathbf{0})
\end{split}
\end{equation} 
 \end{lemma}
For the proof see~\cite{drgona2021stochastic}.

\begin{remark}
Authors in~\cite{montufar2014number} showed that deep \texttt{ReLU} networks can exactly represent any PWA control policy~\eqref{eq:pwa_law} representing the explicit solution of the parametric programming problem~\eqref{eq:param_prog}.
Lemma~\ref{lem:lpv} can be used to generalize this statement as follows. For fully connected neural network with any piecewise linear activations such as \texttt{LeakyReLU}, \texttt{PReLU}, or bendable linear units \texttt{BLU} there exists an equivalent PWA map~\eqref{eq:lpv_dnn}, where the local affine dynamics can be obtained by construction as defined by Lemma~\ref{lem:lpv}.
\end{remark}

Now, by leveraging  Lemma~\ref{lem:lpv} it is straightforward to see that the closed-loop system~\eqref{eq:closed_loop} can be equivalently reformulated as a following PWA system:
\begin{equation}
\label{eq:closed_loop_feedback}
    {\bf x}_{k+1} = ({\bf A}  + {\bf B } \mathbf{H}_{\boldsymbol{\pi}}(\mathbf{x}_k) ) {\bf x}_k + {{\bf B }} \mathbf{b}_{\boldsymbol{\pi}}(\mathbf{x}_k)
\end{equation}
with ${\bf A}  + {\bf B } \mathbf{H}_{\boldsymbol{\pi}}(\mathbf{x}_k) \in \mathbb{R}^{n_x \times n_x}$ representing local linear dynamics of the closed-loop system.

\subsubsection{Contractive neural closed-loop systems}
\label{sec:stability_dnn}

In this paper, we leverage local Lipschitz constants and Banach fixed point theorem to provide sufficient stability conditions for closed-loop  systems~\eqref{eq:closed_loop}.

\begin{theorem}
\label{thm:stability}
A closed-loop system~\eqref{eq:closed_loop} with linear system dynamics and full state feedback policy parametrized by neural network  $\pi_{ {\bf W}}(\mathbf{x})$~\eqref{eq:dnn}
 with piece-wise linear activation functions is globally stable if:
\begin{equation}
  \label{eq:stability_condition}
||{\bf A}  + {\bf B } \mathbf{H}_{\boldsymbol{\pi}}(\mathbf{x}) ||_2 + \frac{||{{\bf B }} \mathbf{b}_{\boldsymbol{\pi}}(\mathbf{x})||_2}{||\mathbf{x}||_2} < 1, \ \forall \mathbf{x} \in {\mathcal{R}_i} \ \forall i \in \mathbb{N}_1^{n_{\mathcal{R}}}
\end{equation}
where $n_{\mathcal{R}}$ represents the total number of distinct linear regions $\mathcal{R}_i$ of a deep neural network~\eqref{eq:dnn}, and $|| \cdot||_2 $ denotes the operator norm.
Or in words, if the local linear dynamics are contractive in all regions of the closed-loop system in its PWA form~\eqref{eq:closed_loop_feedback}, then the system~\eqref{eq:closed_loop} is globally stable.
\end{theorem}

\begin{proof}
The proof is based on the equivalence of the closed-loop system~\eqref{eq:closed_loop} with a PWA map~\eqref{eq:closed_loop_feedback} through Lemma~\ref{lem:lpv}, that holds for all neural control policies~\eqref{eq:dnn} with piece-wise linear activation functions (e.g., \texttt{ReLU}).
Next by leveraging sub-additivity of the operator norms we obtain the following inequalities:
\begin{subequations}
  \label{eq:lipschitz}
  \begin{align}
  || {\bf A} {\bf x} + {\bf B } \pi_{ {\bf W}}({\bf x}) ||_2  &  \le
       || ({\bf A}  + {\bf B } \mathbf{H}_{\boldsymbol{\pi}}(\mathbf{x}) ) {\bf x}||_2  + || {{\bf B }} \mathbf{b}_{\boldsymbol{\pi}}(\mathbf{x})||_2 \\
 \frac{ || {\bf A} {\bf x} + {\bf B } \pi_{ {\bf W}}({\bf x}) ||_2 }{||\mathbf{x}||_2} & \le 
  ||{\bf A}  + {\bf B } \mathbf{H}_{\boldsymbol{\pi}}(\mathbf{x}) ||_2 + \frac{||{{\bf B }} \mathbf{b}_{\boldsymbol{\pi}}(\mathbf{x})||_2}{||\mathbf{x}||_2}   \label{eq:lipschitz_2}
  \end{align}
\end{subequations}
Now it is clear that the right hand side of~\eqref{eq:lipschitz_2} represents local Lipschitz constant $\mathcal{K}(\mathbf{x})$ of the closed-loop system~\eqref{eq:closed_loop}.
Then the condition~\eqref{eq:stability_condition} 
implies Lipschitz continuity with constant $\mathcal{K}(\mathbf{x}) < 1, \ \forall \mathbf{x} \in {\mathcal{R}_i} \ \forall i \in \mathbb{N}_1^{n_{\mathcal{R}}}$, which in turn implies that the closed-loop system~\eqref{eq:closed_loop} is globally contractive.
Now recalling the Banach fixed-point theorem, which states that every contraction converges towards single point equilibirum, we prove the sufficiency of~\eqref{eq:stability_condition}
for a closed-loop stability of the system~\eqref{eq:closed_loop}.
For the proof of the contraction condition for affine maps see~\cite{affine_contract_2006}.
\qed
\end{proof}

% PWA maps are defined by a set of polytopic regions  $\mathcal{R}_i$, each of which contains a local affine map $  \mathbf{A}_{i} + \mathbf{b}_{i}, \ \text{if} \ \mathbf{x} \in {\mathcal{R}_i}$.
%  It is known, that if  $ ||\mathbf{A}_{i}||_2 < 1$ holds, then the affine map is contractive.

% The local linear dynamics of the closed-loop system~\eqref{eq:closed_loop_feedback} is given by a parametric matrix ${\bf A}  + {\bf B } \mathbf{H}_{\boldsymbol{\pi}}(\mathbf{x})$ varying with  the local linear maps of the control policy 
% $\mathbf{H}_{\boldsymbol{\pi}}(\mathbf{x}), \ \text{if} \ \mathbf{x} \in {\mathcal{R}_i}$, parametrized via~\eqref{eq:dnn_LPV}.

\begin{remark}
It is straightforward to see that the assumption~\ref{assume:terminal_set} on local Lyapunov function is satisfied
if the contraction condition~\eqref{eq:stability_condition} holds for all states in the terminal set $\mathcal{X}_f$.
Thus we can use the condition~\eqref{eq:stability_condition} as a metric to 
verify the assumption~\ref{assume:terminal_set} in practice.
\end{remark}

\begin{remark}
Besides the stability condition~\eqref{eq:stability_condition}, the PWA reformulation of the closed-loop system~\eqref{eq:closed_loop_feedback} 
 allows us to apply a wide variety stability analysis methods developed for PWA systems~\cite{PETSAGKOURAKIS2020,IERVOLINO201722,Lin2009,Mignone2000}. Thus due to the generic form of~\eqref{eq:closed_loop_feedback} each of the references methods can be used for stability verification of the policies trained via the proposed DPC policy optimization method.
\end{remark}

\subsection{Connections with MPC Stability Theory}

In this section, we provide an intuitive perspective on the connection between the proposed DPC method and classical MPC stability theory.
Most of the stability analysis methods of MPC
are based on enforcing MPC's value function to be a Lyapunov function~\cite{MAYNE2000789}. 
For further details on the Lyapunov stability of MPC see e.g.~\cite{MAYNE2000789,Zheng1995} and references therein.
 The second class of methods deals with enforcing
contraction constraints on normed state variables~\cite{BFb0109870}.
The structural equivalence of the DPC problem with the classical MPC problem justifies the use of stability principles of MPC in the context of DPC policy optimization via modifying the problem formulation~\eqref{eq:DPC}.

\begin{assumption}
\label{assume:optimality}
The DPC policy gradient optimization algorithms~\ref{algo:DPC_optim}, and~\ref{algo:aDPC_optim}, respectively, converge sufficiently close to a global optimum of the problem~\eqref{eq:DPC}.
\end{assumption}
The strong Assumption~\ref{assume:optimality} is supported by  recent studies in the deep learning literature that show remarkable convergence properties of the stochastic gradient descent (SGD) algorithms.
In particular, it is known that SGD is
able to train complex deep neural networks towards zero training loss while avoiding local minima, thus implying that it can converge towards global optima for these highly non-convex problems~\cite{Jin0NKJ17,zhou2019sgd,Kleinberg2018,Daneshmand2018}.
Moreover, from the control perspective, it has been shown that policy gradient methods such as DPC provide unbiased estimates of the gradient with respect to the policy parameters and are
guaranteed to converge to local optima~\cite{NIPS1999_464d828b}.
Therefore if Assumption~\ref{assume:optimality} holds, then for a trained  control policy ${\bf u}_k^i = \pi_{ {\bf W}}(\boldsymbol \xi_k^i) $, and a distribution  of control parameters $\boldsymbol \xi_k^i \in \Xi \subset \mathbb{R}^n$,
the  parametric Karush–Kuhn–Tucker (KKT) ${\bf g}_{\text{KKT}}({{\bf u}^{\star}}^{i},{\boldsymbol\lambda^{\star}}^{i},{\boldsymbol\mu^{\star}}^{i}) \le {\bf 0}$ of the sampled instances of the associated $i$-th MPC problem~\eqref{eq:mpc_example} 
hold within small violation tolerance $\epsilon$, i.e.,  $||\texttt{ReLU}({\bf g}_{\text{KKT}}({{\bf u}^{\star}}^{i},{\boldsymbol\lambda^{\star}}^{i},{\boldsymbol\mu^{\star}}^{i}))|| \le \epsilon$. 
Thus under the Assumption~\ref{assume:optimality},  with small margin of error, the MPC stability theory applies to DPC policy optimization Algorithms~\ref{algo:DPC_optim}, and~\ref{algo:aDPC_optim}, respectively. This motivates us to explore the use of the following MPC methods to promote the stability of the DPC policies. 

\subsubsection{Terminal Constraints}
This method is based on including additional terminal constraints
${\bf x}_N \in \mathcal{X}_f$ to the DPC problem~\eqref{eq:DPC}.
Stability guarantees~\cite{MAYNE2000789} exist for cases with terminal equality $\mathcal{X}_f = {\bf 0}$ or inequality given as a set $\mathcal{X}_f \in \mathbb{R}^n$ containing the origin.
In DPC we can treat \textit{terminal constraints} 
as any other constraints and enforce them via penalty methods~\eqref{eq:ReLU_ineq}.

\subsubsection{Terminal Penalties}
An alternative approach that can be used in conjunction with terminal constraints is the use of terminal penalties $F({\bf x}_N)$.
Authors in the MPC literature have long before proposed the use of quadratic terminal penalties 
$F({\bf x}_N) = \frac{1}{2} {\bf x}_N {\bf P} {\bf x}_N $, where ${\bf P}$ can be obtained e.g. by solving the Riccati equation, or as a value function of the stabilizing controller~\cite{MAYNE2000789}.
More recent approaches enforce $F({\bf x}_N)$ to be a control Lyapunov function~\cite{Kang2009,WU2019108508,Maaz2014}. 
Again, we can enforce these types of terminal penalties in DPC formulation as extra terms in the loss function~\eqref{eq:policy_loss}.

\subsubsection{Contraction Constraints}
This idea is based on enforcing the contraction of state variables w.r.t. some norm~\cite{BFb0109870}.
The basic form of the contraction constraints stabilizing the states towards the origin is given as:
\begin{equation}
\label{eq:state_contract_con}
    || {\bf x}_{k+1} ||_p \le \alpha || {\bf x}_{k} ||_p, \ \alpha < 1
\end{equation}
Now it is straightforward to see that the contraction can be enforced by penalizing the
constraint~\eqref{eq:state_contract_con} directly in the DPC loss function~\eqref{eq:policy_loss}
again using penalty functions.

\section{Numerical Case Studies}
This section presents simulation results on five example systems demonstrating the capabilities of the DPC policy optimization algorithm to i) learn to stabilize unstable linear systems, 
ii) learn to control systems with multiple inputs and outputs, 
iii) satisfy state and input constraints, 
iv) adaptively update the parameters of the prediction model and  control policy online, to safely control an unknown linear system affected by unobserved parametric and additive uncertainties.
Furthermore, we demonstrate superior sampling efficiency compared to model-free RL algorithms, faster online execution compared to implicit MPC, and smaller memory footprint and better scalability compared to explicit MPC.
The presented experiments are implemented in Pytorch-based toolbox NeuroMANCER~\cite{Neuromancer2021} and are available on Github\footnote{\url{https://github.com/pnnl/deps_arXiv2020}}.

\subsection{Example 1: Stabilizing Unstable Double Integrator}
 \label{sec:stabilization2}
 
Here we demonstrate 
the capabilities of DPC to learn  a stabilizing neural feedback policy for an unstable system with known dynamics. We control the unstable double integrator: 
 \begin{equation}
 \label{eq:double_int}
{\bf x}_{k+1} = \begin{bmatrix} 1.2 & 1.0 \\ 0.0 & 1.0\end{bmatrix}  {\bf x}_k + \begin{bmatrix} 1.0  \\ 0.5  \end{bmatrix}   {\bf u}_k
 \end{equation}
Subject to state and input constraints given as 
 ${\bf x}_k \in \mathcal{X} := \{ {\bf x} \mid -10 \le { x} \le 10 \}$, and ${\bf u}_k \in\mathcal{U} := \{ {\bf u}  \mid -1 \le { u} \le 1  \}$, respectively.
For control we consider the following objective.
 \begin{equation}
 \label{eq:empc_qp}
   \mathcal{L}_{\texttt{MPC}} =  \sum_{k=0}^{N-1}  \big(
 || {\bf x}_k||_{Q_x}^2  + || {\bf u}_k ||_{Q_u}^2 \big)
 \end{equation}
 
  As a benchmark we synthesize  an explicit MPC policy  solved with classical multi-parametric programming solver using MPT3 toolbox~\cite{MPT3:2013} with prediction horizon of $N = 10$, and weights  $Q_x = 1$, $Q_u = 1$,
 for whose the surface is shown on the left in Fig.~\ref{fig:e1:policies}.
  
 \subsubsection{Training}
\begin{figure}
    \centering
    % \hspace{-0.3cm}
    % \includegraphics[width=.35\textwidth]{./figs/empc_policy.pdf} 
    % \includegraphics[width=.35\textwidth]{./figs/dpc_policy.pdf}
    \includegraphics[width=.48\textwidth]{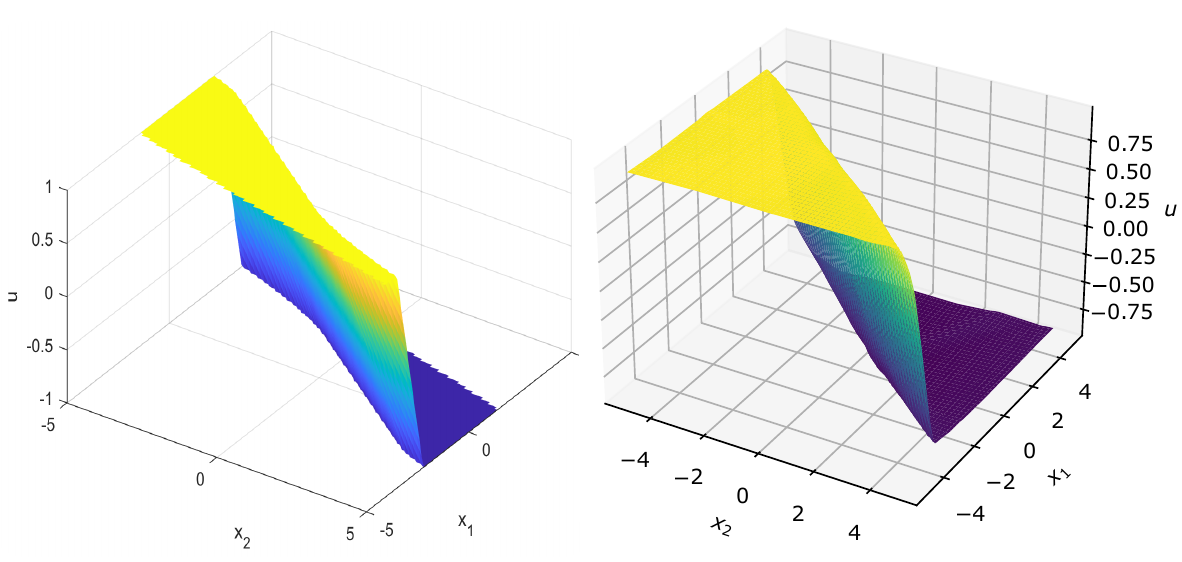}
    \caption{Surfaces of PWA explicit MPC policy (left)  and neural DPC policy (right) for  double integrator system~\eqref{eq:double_int}.}
    \label{fig:e1:policies}
\end{figure}

 In the case of DPC, we learn a full state feedback neural policy
 ${\bf u}_k = \boldsymbol \pi_{{\bf W}}({\bf x}_k)$
 in the closed-loop system~\eqref{eq:closed_loop} via  Algorithm~\ref{algo:DPC_optim} for whose the surface is shown on the right in Fig.~\ref{fig:e1:policies}.
 A side by side comparison in Fig.~\ref{fig:e1:policies} shows almost identical control surfaces of compared policies. 
 For the DPC policy training we used the MPC loss function~\eqref{eq:empc_qp} subject to state  ${\bf x}_k \in \mathcal{X}$, input ${\bf u}_k \in \mathcal{U}$, as well as 
 terminal set constraints ${\bf x}_N \in \mathcal{X}_f :=  \{ {\bf x} \mid -0.1 \le { x} \le 0.1 \}$. All constraints are implemented using penalties~\eqref{eq:ReLU_ineq}  with  weights $Q_{h} = 10$, $Q_{g} = 100$, $Q_{\mathcal{X}_f} = 1$, while for the control objective~\eqref{eq:empc_qp} 
 we consider prediction horizon $N = 1$, and weights $Q_x = 5$, $Q_u = 0.5$.
 We train the neural policy~\eqref{eq:dnn} $ \pi_{{\bf W}}({\bf x}): \mathbb{R}^2 \to \mathbb{R}$ with $4$ layers, $20$ hidden states, with bias, and \texttt{ReLU} activation functions.
 We use a training set $\mathcal{X}^{\text{train}}$ of $3333$  normally sampled initial conditions ${\bf x}^i_0$  fully covering the admissible set  $\mathcal{X}$.
 
  \subsubsection{Closed-loop perofmance}
\begin{figure}
    \centering
    % \hspace{-0.3cm}
    \includegraphics[width=.40\textwidth]{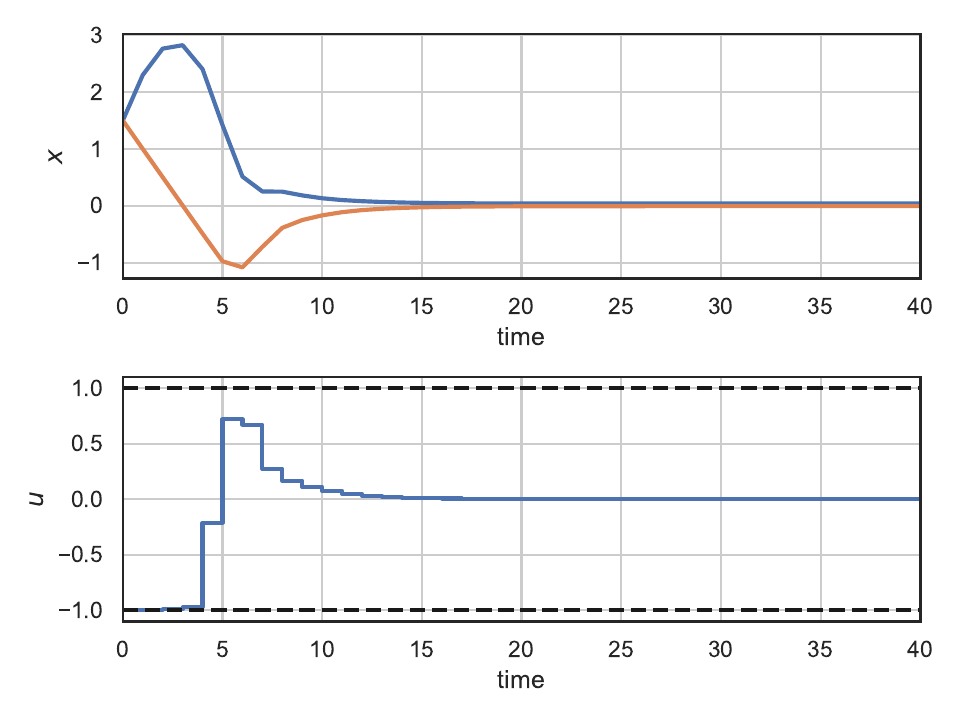}
    \caption{Closed-loop trajectories of  system~\eqref{eq:double_int} controlled by stabilizing neural feedback policy trained using  Algorithm~\ref{algo:DPC_optim} with DPC problem formulation~\eqref{eq:DPC} including terminal constraint ${\bf x}_N \in \mathcal{X}_f$.}
    \label{fig:e1:DPC_cl}
\end{figure}

\begin{figure}
    \centering
    % \hspace{-0.3cm}
    \includegraphics[width=.23\textwidth]{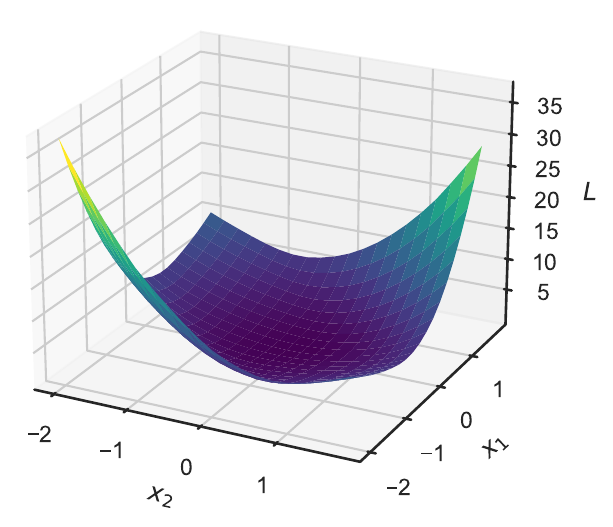}
    \includegraphics[width=.23\textwidth]{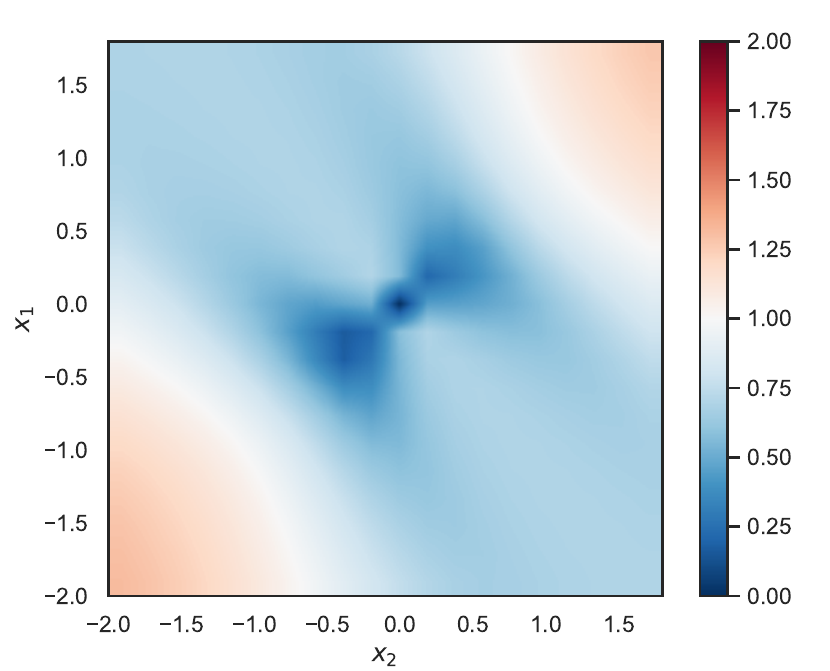}
    \caption{DPC loss function~\eqref{eq:policy_loss} on the left, and on the right, regions with  contractive (blue) and divergent (red) dynamics  plotted over state space of a double integrator system~\eqref{eq:double_int} controlled by stabilizing  neural feedback policy trained using DPC Algorithm~\ref{algo:DPC_optim}.}
    \label{fig:e1:loss}
\end{figure}

 Fig.~\ref{fig:e1:DPC_cl} shows resulting converging trajectories of the closed-loop system dynamics
 controlled by a stabilizing neural policy learned using DPC formulation with terminal constraints.
 Finally on the left hand side of Fig.~\ref{fig:e1:loss} we visualize the DPC loss function~\eqref{eq:policy_loss} (left)
 for a trained policy over a state space to demonstrate
 that it is indeed a Lyapunov function. 
 While on the right of Fig.~\ref{fig:e1:loss} we evaluate the contraction constraint~\eqref{eq:state_contract_con}
 showing state space regions with
  contractive  $\alpha < 1$ (blue) and diverging $\alpha > 1$ (red)  dynamics. 
% 
% \begin{figure}
%     \centering
%     % \hspace{-0.3cm}
%     \includegraphics[width=.40\textwidth]{./figs/closed_loop_dpc_untuned.pdf}
%     \caption{Closed-loop system trajectories of  system~\eqref{eq:double_int} controlled by neural feedback policy trained using  Algorithm~\ref{algo:DPC_optim} with poorly tuned DPC problem~\eqref{eq:DPC} without terminal constraints.}
%     \label{fig:e1:DPC_cl_2}
% \end{figure}
% 

% \add{TODO: add stability regularizaiton during training?
% }
% \add{TODO: check eigenvalues of the policy to see if it is stable in the closed loop. }
% \add{TODO: plot control policies, and closed loop trajectories - different initial conditions?
% }
% \add{TODO: plot feasibility/ stable state space regions}
% \add{TODO: mpc trajectories over DPC in different colors???}
% \add{plot DPC control invariant sets}

\subsection{Example 2: Stabilizing PVTOL Aircraft Model}
 \label{sec:stabilization1}
 
  We consider a planar  vertical take-off and landing (PVTOL) aircraft model as appeared in~\cite{Astromfeedbacksystems}.
 The model is represented as a multiple-input multiple-output (MIMO) linear state space model, where the states  ${\bf x}_k \in \mathbb{R}^6$ represent  positions $(x, y)$,  velocities $(\dot{x}, \dot{y})$, and orientation of the center of mass  ${\theta}$, and its derivative  $\dot{\theta}$. The control inputs ${\bf u} \in \mathbb{R}^2$ represent a pair of forces generated by the aircraft thrusters. For the control purposes, the continuous-time PVTOL model is discretized with a sampling rate of $0.2$ seconds.
 
 \subsubsection{Training}
To train the control policy via DPC Algorithm~\ref{algo:DPC_optim} we sample $9000$  normally distributed initial conditions, where we use $3000$ samples for train, validation, and test set, respectively.
The DPC problem is formulated with quadratic control objective~\eqref{eq:empc_qp} with prediction horizon $N =10$, and weights $Q_x = 3$, $Q_u = 0.1$,
 subject to state and input constraints given as 
 ${\bf x}_k \in \mathcal{X} := \{ {\bf x} \mid -5 \le { x} \le 5 \}$, and ${\bf u}_k \in\mathcal{U} := \{ {\bf u} \mid -5 \le { u} \le 5 \}$, with constraints penalty weights $Q_{h} = 2,$ $Q_{g} = 2$, respectively.
  Again, we train the neural policy~\eqref{eq:dnn} $ \pi_{{\bf W}}({\bf x}): \mathbb{R}^6 \to \mathbb{R}^{N \times 2}$ with $4$ layers, $20$ hidden states, with bias, and \texttt{ReLU} activation functions.
In the closed-loop the DPC policy is implemented in receding horizon (RHC) thus is being reduced to a mapping $ \pi_{{\bf W}}({\bf x}): \mathbb{R}^6 \to \mathbb{R}^{2}$.

\subsubsection{Closed-loop performance}
\begin{figure}[ht!]
\centering
\begin{subfigure}{0.48\textwidth}
\centering
    \includegraphics[width=0.9\linewidth]{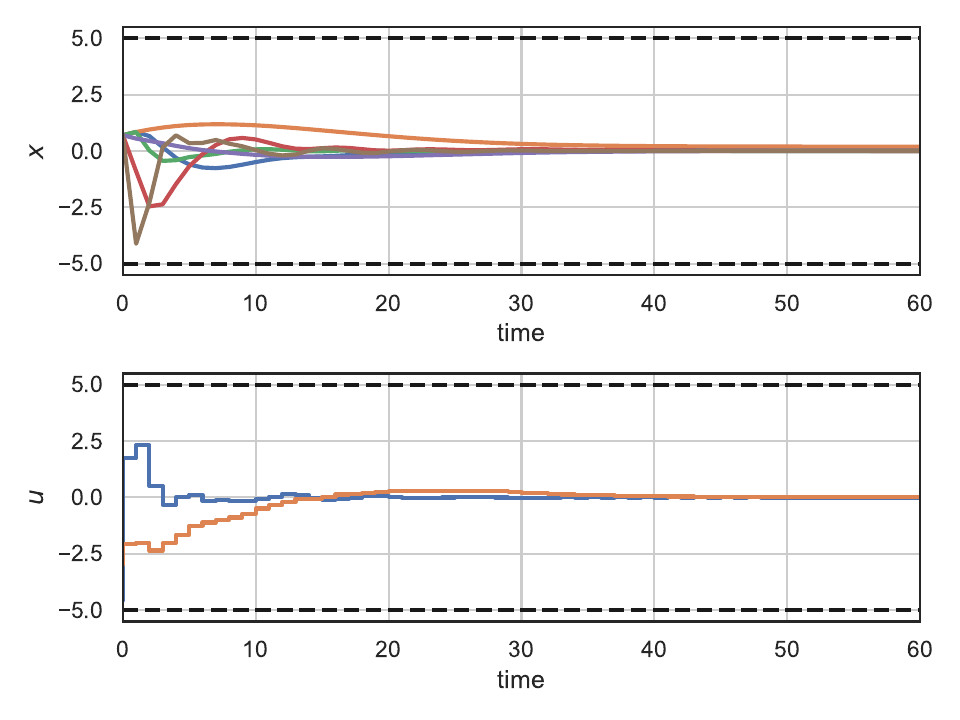}
        \vspace{-0.3cm}
    \caption{DPC}
    \label{fig:vtol_cl1:dpc}
\end{subfigure} \\
\begin{subfigure}{0.48\textwidth}
\centering
    \includegraphics[width=0.9\linewidth]{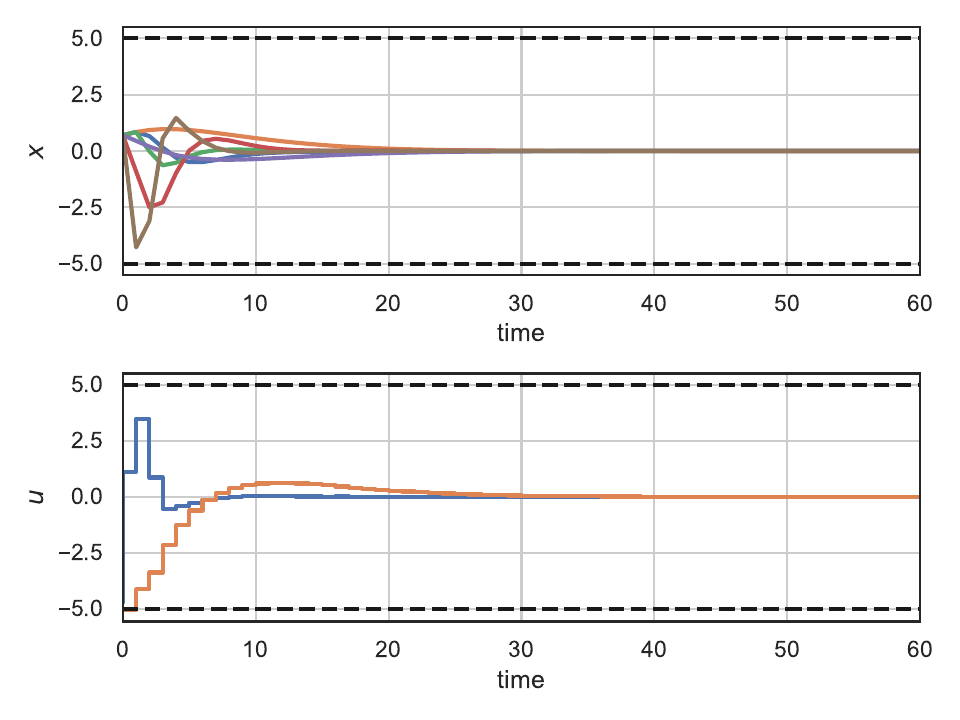}
    \vspace{-0.3cm}
    \caption{approximate MPC}
    \label{fig:vtol_cl1:ampc}
\end{subfigure}
  \caption{Closed-loop trajectories of the PVTOL aircraft model controlled by stabilizing DPC (top) and approximate MPC (bottom) policies. }
\label{fig:vtol_cl1}
\end{figure}

In Fig.~\ref{fig:vtol_cl1} we show the comparison of the  DPC policy optimization against approximate MPC (aMPC) trained in supervision on control inputs examples generated by solving the original MPC problem~\eqref{eq:mpc_example} via a quadratic programming solver. We keep the policy architectures, prediction horizon, and number of training samples identical for a fair comparison.
We demonstrate that the neural DPC policy trained offline via Algorithm~\ref{algo:DPC_optim} can stabilize the closed-loop system dynamics from different initial conditions, while satisfying state and input constraints.
In comparison with aMPC, the trained DPC policies take more conservative control actions, keeping a larger margin from the control constraints and results in a slightly higher settling time. 
This  behavior can be explained by a full batch training that optimizes the DPC policy w.r.t. an average performance~\eqref{eq:DPC:objective} on the whole training dataset distribution. 
On the other hand, the aMPC is trained on samples generated by deterministic MPC that operates close to the constraints. 

% \begin{figure}
%     \centering
%     % \hspace{-0.3cm}
%     \includegraphics[width=.40\textwidth]{./figs/pvtol_dpc_cl1.pdf}
%         \includegraphics[width=.40\textwidth]{./figs/pvtol_dpc_cl2.pdf}
%     \caption{Closed-loop trajectories of the PVTOL aircraft model controlled by stabilizing DPC policy from different initial conditions. The neural policy was trained offline using  Algorithm~\ref{algo:DPC_optim}.}
%     \label{fig:vtol_cl1}
% \end{figure}

\subsection{Example 3: Tracking Control of a Quadcopter Model}
 \label{sec:ex3_quadcopter}

In this example, we demonstrate  scalability of DPC to larger systems 
using a linear quadcopter model\footnote{\url{https://osqp.org/docs/examples/mpc.html}} with
twelve states ${\bf x}_k \in \mathbb{R}^{12}$, and four control inputs  ${\bf u} \in \mathbb{R}^4$.
Our objective is to track the reference with the $3$-rd state representing the controlled output ${\bf y}_k \in \mathbb{R}$, and  stabilize rest of the states, while satisfying state and input constraints.

 \subsubsection{Training}
To optimize the policy, we use Algorithm~\ref{algo:DPC_optim} with $30,000$ samples of feasible normally distributed initial conditions, where we use $10,000$ samples for train, validation, and test set, respectively.
We consider the DPC problem with quadratic control objective:
\begin{equation}
 \label{eq:reference}
   \mathcal{L}_{\texttt{MPC}} =  \sum_{k=0}^{N-1}  \big(
 || {\bf y}_k - {\bf r}_k||_{Q_r}^2  + 
  || {\bf x}_k ||_{Q_x}^2
 \big)
 \end{equation}
using prediction horizon $N =10$, and weights $Q_r = 20$, $Q_x = 5$.
The system is subject to 
 state and input constraints 
 ${\bf x}_k \in \mathcal{X} := \{ {\bf x} \mid -10 \le { x} \le 10 \}$, and ${\bf u}_k \in\mathcal{U} := \{ {\bf u} \mid -1.0 \le { u} \le 2.5 \}$, with constraints penalty weights $Q_{h} = 1,$ $Q_{g} = 2$, respectively.
To promote the stability of the closed-loop system during training, we consider the contraction constraint~\eqref{eq:state_contract_con} with scaling factor $\alpha = 0.8$ and penalty weight $Q_{c} = 1$.
We train the full state feedback neural policy~\eqref{eq:dnn} $ \pi_{{\bf W}}({\bf x}): \mathbb{R}^{12} \to \mathbb{R}^{N \times 4}$ with $2$ layers, $100$ hidden states, and \texttt{ReLU} activation functions.
In the closed-loop simulations, we again use 
RHC to implement only the first predicted control action.

\subsubsection{Closed-loop performance}
\begin{figure}
\centering
\begin{subfigure}{0.48\textwidth}
\centering
    \includegraphics[width=0.9\linewidth]{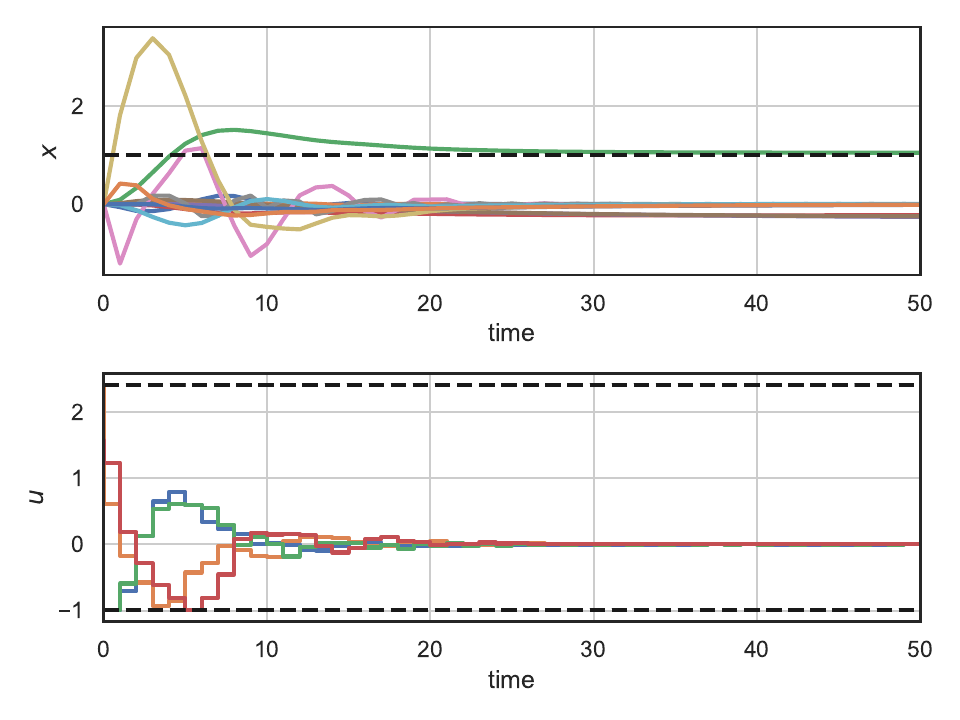}
        % \vspace{-0.3cm}
    % \caption{DPC}
    \label{fig:quadcopter:dpc}
\end{subfigure} \\
\begin{subfigure}{0.48\textwidth}
\centering
    \includegraphics[width=0.9\linewidth]{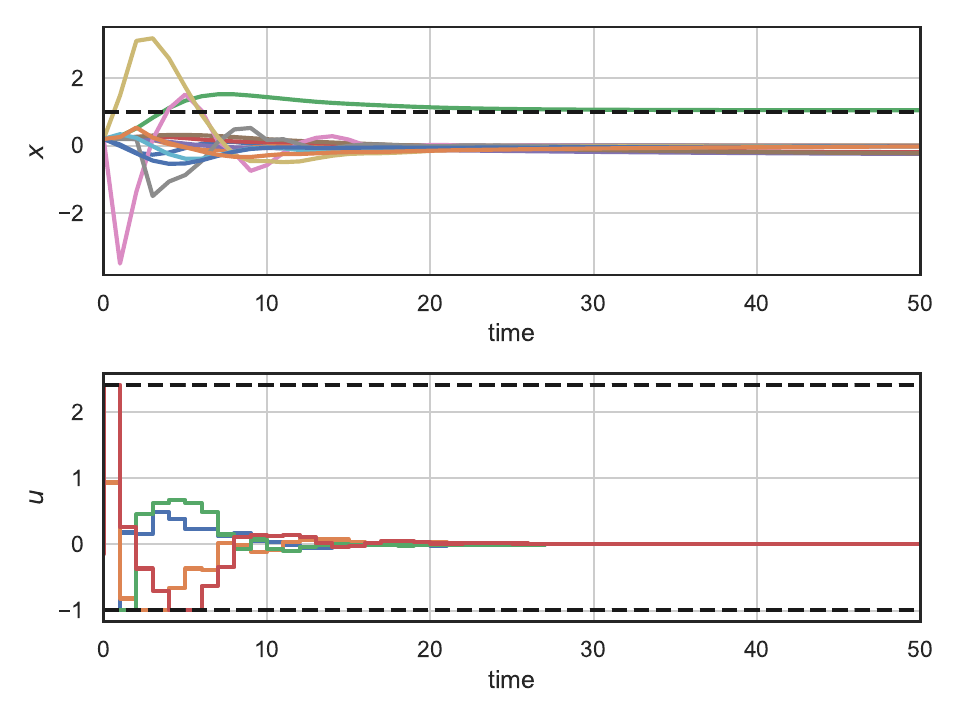}
    % \vspace{-0.3cm}
    % \caption{Initial condition 2.}
    \label{fig:quadcopter:ampc}
\end{subfigure}
    \caption{Closed-loop trajectories of the quadcopter system starting from two different initial conditions controlled by DPC policy trained via  Algorithm~\ref{algo:DPC_optim}  using control objective~\eqref{eq:reference} and constraints penalties~\eqref{eq:ReLU_ineq} including contraction constraint~\eqref{eq:state_contract_con}.}
    \label{fig:quadcopter}
\end{figure}

Fig.~\ref{fig:quadcopter} plots the closed-loop control performance of the quadcopter model from two different initial conditions controlled with the trained full state feedback neural policy.
We demonstrate that the trained DPC policy can 
simultaneously track the desired reference, and stabilize the selected states while satisfying state and input constraints. 
From the computational perspective, in Table~\ref{tab:cpu_quadcopter} we compare the online computational time associated with the evaluation of the trained DPC policy compared against the implicit MPC in the quadratic form implemented in CVXPY~\cite{diamond2016cvxpy} and solved with the OSQP solver~\cite{osqp2020}.
We report on average almost two orders of magnitude speedup in the maximum and mean evaluation time compared to the online optimization solver.
Due to the large state dimensions, the presented problem is not feasible with traditional parametric programming solvers. 
Thus by solving this problem, we demonstrate the scalability of the proposed DPC method beyond the limitations of classical explicit MPC while providing significant speedups compared to the online state-of-the-art convex optimization solver.
\begin{table}[ht!]
    \centering
       \caption{Comparison of online computational time of the proposed DPC policy  against implicit MPC solved with OSQP.}
\begin{tabular}{lll}
\toprule
{online evaluation time} &   mean  [$1e^{-3}$ s] & max  [$1e^{-3}$ s]   \\
\midrule
DPC &   0.196 & 0.997 \\
MPC (OSQP) &  8.676 &   95.744  \\
\bottomrule
\end{tabular}
    \label{tab:cpu_quadcopter}
\end{table}

\subsection{Example 4: Parametric Obstacle Avoidance Problem}
 \label{sec:ex_4_obstacle}

In this example, we demonstrate handling parametric nonlinear constraints for obstacle avoidance problem by  learning parametric neural policies
using DPC policy optimization Algorithm~\ref{algo:DPC_optim}.
We consider the  double integrator system:
\begin{equation}
 \label{eq:double_int}
{\bf x}_{k+1} = \begin{bmatrix} 1.0 & 0.1 \\ 0.0 & 1.0\end{bmatrix}  {\bf x}_k + \begin{bmatrix} 1.0 & 0.0 \\ 0.0 & 1.0  \end{bmatrix}   {\bf u}_k
 \end{equation}
 Where ${\bf x} \in \mathbb{R}^2$ and ${\bf u} \in \mathbb{R}^2$ are states and inputs that are subject to the  box constraints 
 ${\bf x}_k \in \mathcal{X} := \{ {\bf x} \mid -10 \le { x} \le 10 \}$, and ${\bf u}_k \in\mathcal{U} := \{ {\bf u}  \mid -1 \le { u} \le 1  \}$, respectively.
Additionally, we consider  parametric nonlinear constraints representing an obstacle in the state space:
\begin{equation}
 \label{eq:obstacle}
p^2 \le b ({\bf x}_{1,k} - c)^ 2 + ({\bf x}_{2,k} - d) ^ 2
 \end{equation}
where 
${\bf x}_{i,k}$ denotes the $i$-th system dynamics state, and
$p$, $b$, $c$, $d$ are scalar-valued parameters defining the volume, shape, and center of the eliptic obstacle defined by equation~\eqref{eq:obstacle}. 

 \subsubsection{Training}
Again we use the DPC policy optimization Algorithm~\ref{algo:DPC_optim} for training the  parametric neural policy 
$ \pi_{{\bf W}}({\bf x}, \boldsymbol \xi): \mathbb{R}^{7} \to \mathbb{R}^{N \times 2}$
with $4$-layer, each having $100$ hidden neurons, and \texttt{ReLU} activation functions, where $\boldsymbol \xi = [b, c, d, {\bf r}_N ]$.
In the problem formulation~\eqref{eq:DPC}, we consider the following control objective
with the prediction horizon $N=20$:
\begin{equation}
 \label{eq:obstacle_loss}
 \begin{split}
   \mathcal{L}_{\texttt{MPC}} =  
   || {\bf x}_N - {\bf r}_N||_{Q_r}^2  +   \sum_{k=0}^{N-2} || {\bf u}_{k+1} - {\bf u}_k ||_{Q_{du}}^2 +\\
  \sum_{k=0}^{N-1}  \big(
  || {\bf x}_{k+1} - {\bf x}_k ||_{Q_{dx}}^2 + 
  || {\bf u}_k ||_{Q_{u}}^2
 \big) 
 \end{split}
 \end{equation}
where the first term penalizes the terminal state condition with parametric reference ${\bf r}_N$,
the second term penalizes the energy used in control actions, while the third and fourth terms represent control action and state smoothing penalties.
For scaling the control objective terms we use following weight factor values $Q_r = 1.0$, $Q_{dx} =1.0$, $Q_{du} = 10.0$,  $Q_u = 10.0$, and we use $Q_h = 100.0$ for state constraint penalties that include the obstacle avoidance constraint~\eqref{eq:obstacle} and box constraints on states and control actions.
For  datasets we sample $30000$  uniformly distributed initial state conditions ${\bf x}_{0}^i$, constraints parameters  $p^i$, $b^i$, $c^i$, $d^i$, and terminal state references ${\bf r}_N^i$, and use one third of data for train, validation, and test set, respectively.

 \subsubsection{Optimal control performance}
\begin{figure}[ht!]
\centering
% \centering
    \includegraphics[width=0.49\linewidth]{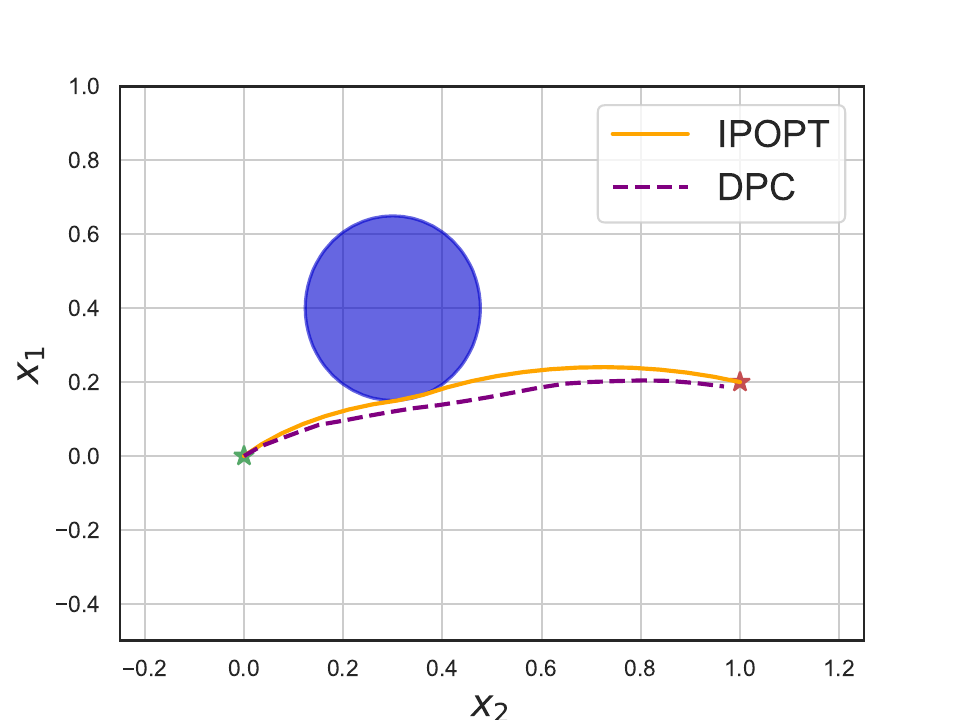}
        \includegraphics[width=0.49\linewidth]{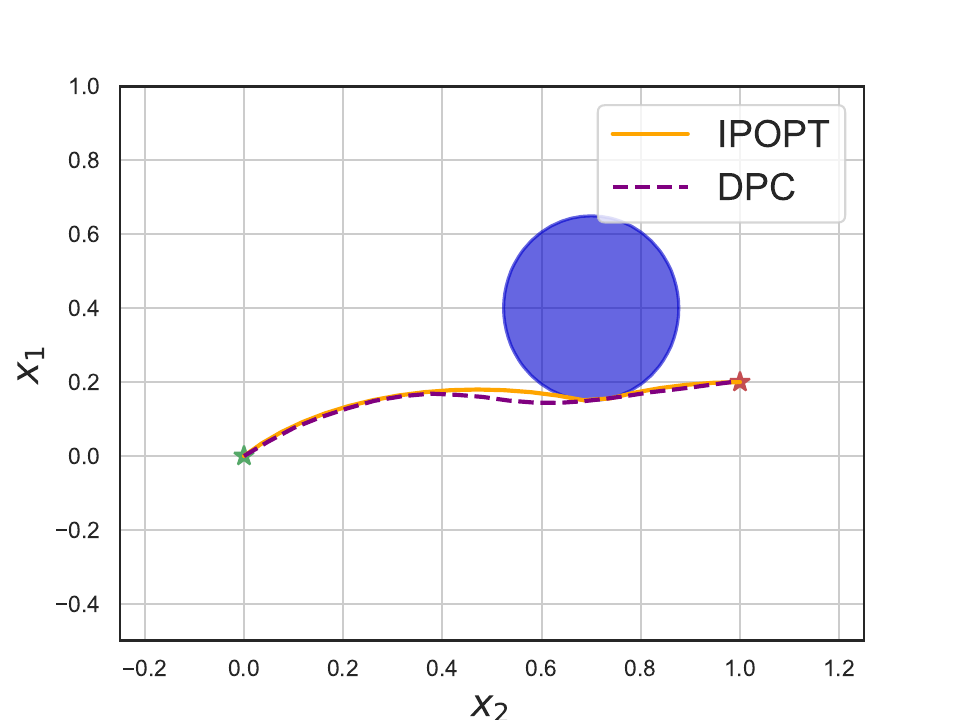}
    \includegraphics[width=0.49\linewidth]{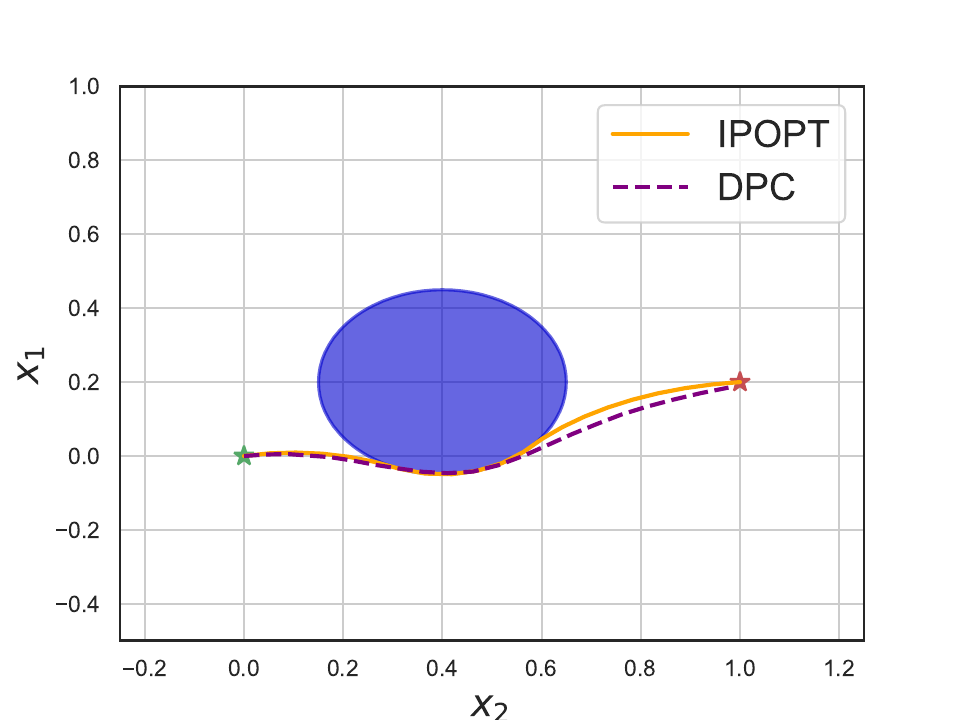}
    \includegraphics[width=0.49\linewidth]{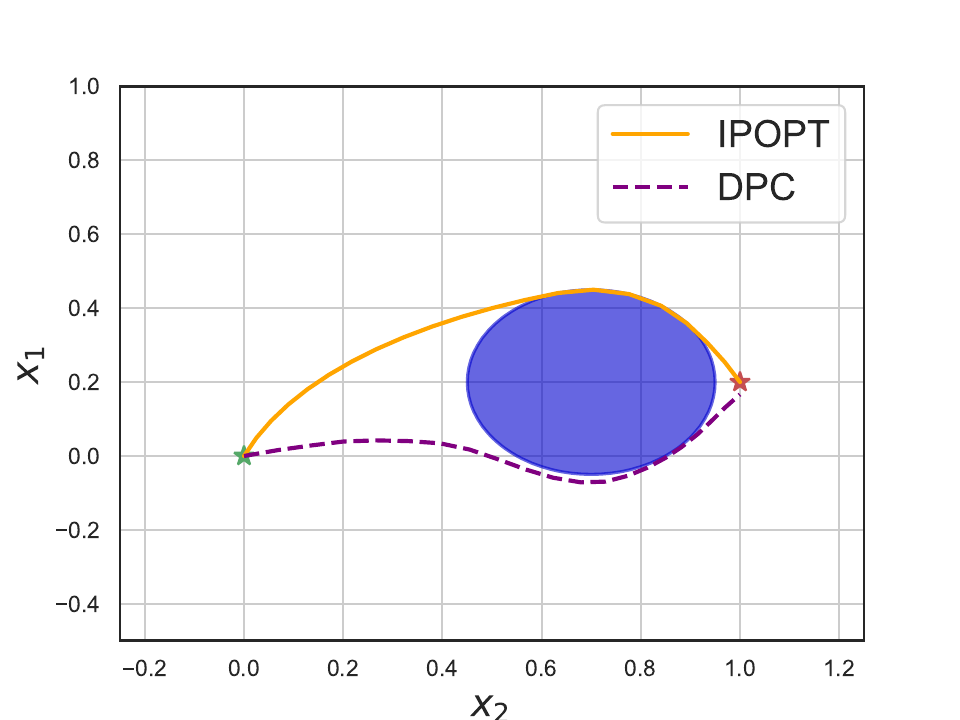}
    \caption{Trajectories in different parametric scenarios of the obstacle avoidance problem~\eqref{eq:obstacle}  computed online using IPOPT and offline using DPC policy optimization Algorithm~\ref{algo:DPC_optim}.}
    \label{fig:obstacle}
\end{figure}

The obstacle constraint~\eqref{eq:obstacle} casts the overall parametric optimal control problem~\eqref{eq:DPC} to be nonlinear. Hence for the comparison, we implement the implicit MPC~\eqref{eq:mpc_example} with nonlinear constraints in the CasADi framework~\cite{Andersson2019} using the nonlinear programming solver IPOPT~\cite{Wchter2006OnTI}.
In Fig.~\ref{fig:obstacle} we visualize, for different parametric scenarios, the trajectories computed online using IPOPT and those computed offline using our DPC policy optimization Algorithm~\ref{algo:DPC_optim}.
We demonstrate that DPC can generate near-optimal trajectories that satisfy a distribution of parametric nonlinear constraints while being more computationally efficient in online time than the state-of-the-art IPOPT solver.  
In particular, Table~\ref{tab:cpu_obstacle} compares mean and maximum online computational time associated with the evaluation of the learned DPC policy 
 with implicit nonlinear MPC solved using IPOPT. 
 On average, the explicit neural DPC policy provides almost  $40$-times speedup and $30$-times speedup in the worst-case compared to IPOPT online solver without sacrificing the control performance on the presented obstacle avoidance problem.
\begin{table}[ht!]
    \centering
       \caption{Comparison of online computational time of the proposed DPC policy and implicit MPC solved with IPOPT.}
\begin{tabular}{lll}
\toprule
{online evaluation time} &   mean  [$1e^{-3}$ s] & max  [$1e^{-3}$ s] \\
\midrule
DPC &   3.492   &   7.001\\
MPC (IPOPT)  &   135.291  &   214.420  \\
\bottomrule
\end{tabular}
    \label{tab:cpu_obstacle}
\end{table}

\subsection{Example 5: Adaptive Control of an Uncertain System}
 \label{sec:dataset}
 
%  \paragraph{Ground truth system dynamics model}
%  \paragraph{Uncertain system dynamics}

In this example we demonstrate adaptive control capabilities together with robustness to parametric and additive uncertainties of the neural policies learned via adaptive DPC Algorithm~\eqref{algo:aDPC_optim}.
We consider the following linear system with varying parametric and additive uncertainties:
\begin{equation}
\label{eq:truth:model:uncertain}
    {\bf x}_{k+1} = \bigr({\bf A} \odot {\bf V}_k \bigl) {\bf x}_k + {\bf B } {\bf u}_k + {\bf E} {\bf d}_k + {\bf w}_k,
\end{equation}
where  $\odot$ is elementwise matrix multiplication.  To assess the robustness of investigated control policies we consider 
 unknown parametric ${\bf V}_k$ and additive ${\bf w}_k$ uncertainties,
with ${\bf V}_k$ elements distributed $\mathcal{N}(1.0,0.01)$, ${\bf w}_k$ elements distributed $\mathcal{N}(0, 0.1)$. 

The model~\eqref{eq:truth:model:uncertain} in our case study represents thermal dynamics of a building~\cite{Drgona:CDC13}.
 The states  ${\bf x}_k \in \mathbb{R}^4$ represent  temperatures, control input ${\bf u} \in \mathbb{R}$ represents heat flow, and ${{\bf d}_k \in \mathbb{R}^3}$ represent measured disturbances, respectively. 
 We run the  simulation of the ground truth model~\eqref{eq:truth:model:uncertain} to obtain total of  $6048$ samples corresponding to three weeks of the building dynamics. The training, validation, and test sets are the $1^{st}$, $2^{nd}$, and $3^{rd}$ weeks of simulation, respectively. Thus in this example, we will demonstrate the DPC's ability to learn safe control policies with small amount of training data using only $2016$ samples.
 The measurable disturbance trajectories  $\mathbf{D}$ are visualized at the bottom of Figure~\ref{fig:SysID_Control:cl}, while
   the reference trajectory  $\mathbf{R}$ for closed-loop control task  is generated as a static sine wave in range \SIrange{18}{22}{\celsius} and period of one day as shown at the top of Figure~\ref{fig:SysID_Control:cl}.

   We evaluate each investigated method on
 reference tracking MSE: $\frac{1}{T} \sum_{k=1}^T ||{\bf r}_{k} - {\bf \tilde{x}}_{k, 4}||^2_2$,
  energy minimization given as  mean absolute (MA) value of the control signal: $ \frac{1}{T} \sum_{k=1}^T  |{\bf u}_{k}|  $, and 
  state constraints violations 
  evaluated as MA value of the slack variables: $ \frac{1}{T} \sum_{k=1}^T  |{\bf s}_{k}^x|  $.

\subsubsection{Offline Policy Optimization with Known Nominal Model}
\label{sec:results:control}

We compare the proposed offline DPC policy optimization against two classical control methods (LQR, LQI), three formulation variants of MPC (nominal, robust, and stochastic), and three deep RL algorithms (PPO2, A2C, ACKTR), respectively.
In this scenario, all of the model-based methods, including DPC, are designed with known nominal matrices of the system~\eqref{eq:truth:model:uncertain}.
The details on control design of each method are elaborated in the Appendix~\ref{sec:results:control}.

\begin{figure}[htb]
    \centering
    \hspace{-0.3cm}
    \includegraphics[width=.44\textwidth]{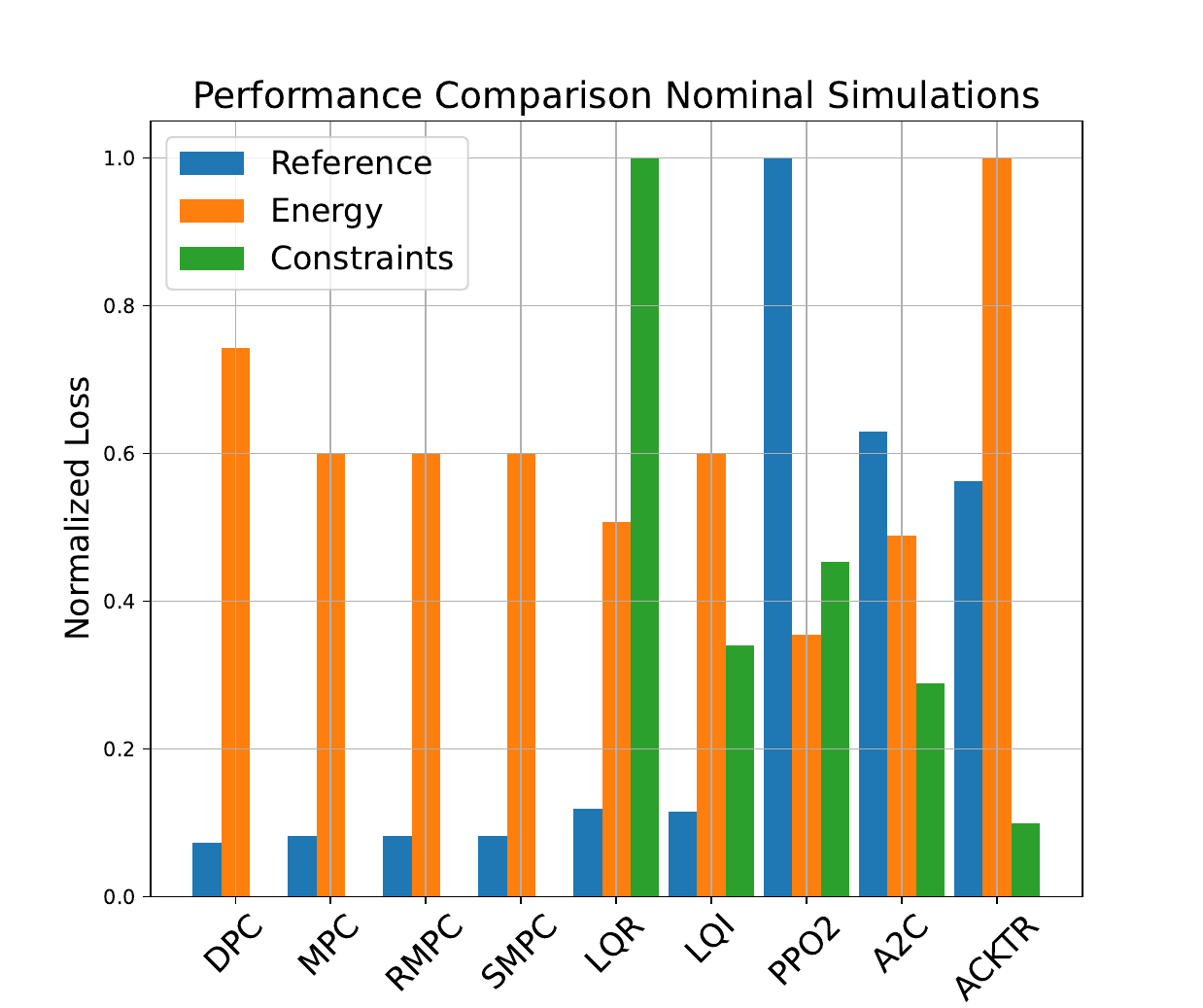}
    \caption{Control performance comparison in simulations without uncertainties.}
    \label{fig:performance:nominal}
\end{figure}

\begin{figure}[htb]
    \centering
    \hspace{-0.3cm}
    \includegraphics[width=.44\textwidth]{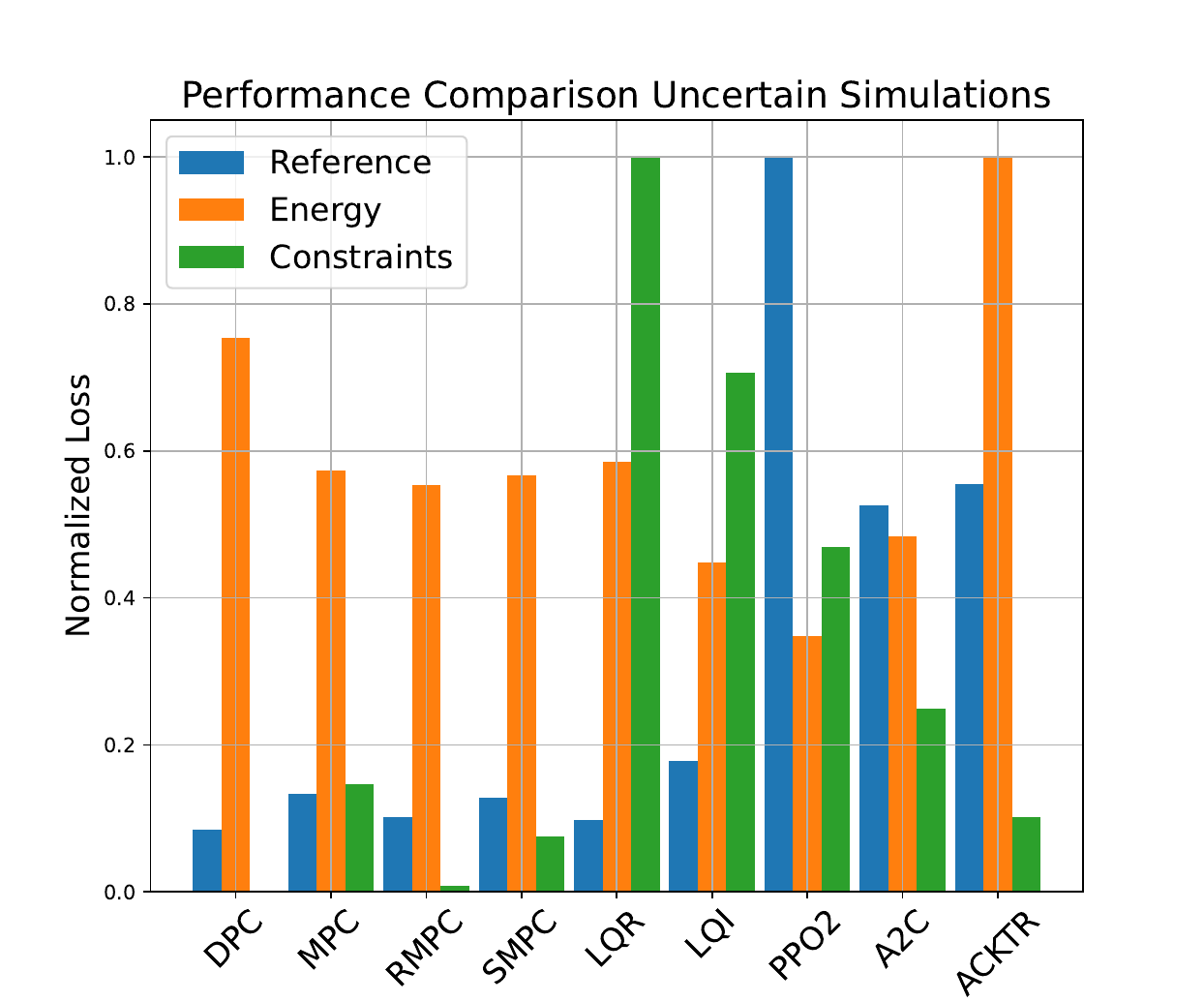}
    \caption{Control performance comparison in simulations with parametric $v_k$ and additive $w_k$ uncertainties.}
    \label{fig:performance:uncertain}
\end{figure}

% performance comparison
Table~\ref{tab:control:GT} shows the corresponding closed-loop control performance of the investigated methods evaluated with and without uncertainties ${\bf v}_k$ and ${\bf w}_k$ acting on the simulation model.
% figures
Figure~\ref{fig:performance:nominal} and Figure~\ref{fig:performance:uncertain} compare visually the control performance of nominal simulations without uncertainties (second column of Table~\ref{tab:control:GT}), and simulations with both parametric and additive uncertainties (fifth column of Table~\ref{tab:control:GT}), respectively.

% table
\begin{table}[htb]
    \centering
       \caption{Control performance on reference tracking, energy use, and constraints violations evaluated on closed-loop simulations with various degrees of uncertainties ${\bf v}_k$ and ${\bf w}_k$. Evaluated controllers were learned/designed with the ground truth model.}
\begin{tabular}{lllll}
\toprule
{Test set} &  No unc.    & ${\bf w}_k$ & ${\bf v}_k$ & ${\bf w}_k$ \& ${\bf v}_k$   \\
\midrule
% \multicolumn{5}{c}{\textbf{MPC without disturbance forecast}}    \\
% MPC MSE ref. &  2.449   &  2.516  &  4.725  &  4.884      \\
% MPC MA ene.  &  980   &   1064   &   853    &     892    \\
% MPC MA con.  &  0.000   &   0.000  &    0.041  &    0.183 \\
% \midrule
\multicolumn{5}{c}{\textbf{Proposed DPC}}    \\
MSE ref.  & \bf{1.244}   &    1.470 &  \bf{1.991}    &    \bf{2.355}     \\
MA ene.  & 1111 &    1092 & 1177 &   1139   \\
MA con.  &  \bf{0.000} &  \bf{0.000}     &   \bf{0.000}    &   \bf{0.000}    \\
 \midrule
\multicolumn{5}{c}{\textbf{Nominal MPC }}    \\
MSE ref. &  1.398   & 1.785  &   3.343 &   3.711     \\
MA ene.  &  \bf{897}  &   \bf{847} &   917 &   866      \\
MA con.  &  \bf{0.000}  &   0.132  &  1.002  &  1.106   \\
\midrule
\multicolumn{5}{c}{\textbf{Robust MPC }}    \\
MSE ref.  &  1.405   &  1.640    &    2.727   &  2.839   \\
MA ene.  &    899    &  892    &     \bf{910}    &    \bf{836}     \\
MA con.  & \bf{0.000}   &  \bf{0.000}   &   0.007      &  0.066      \\
\midrule
\multicolumn{5}{c}{\textbf{Stochastic MPC }}    \\
MSE ref.  &  1.398     &  \bf{1.422} &   2.238     &    3.579    \\
MA ene.  &   \bf{897}     &  995  &    977    &    856    \\
MA con.  &  \bf{0.000}   &  0.098   &  0.476      &  0.572      \\
\midrule
\multicolumn{5}{c}{\textbf{LQR}}    \\
MSE ref. &  2.024  &  2.184   &  2.655 & 2.711  \\
MA ene.  &  758 &    793   &   880     &   883     \\
MA con.  &  5.574  &  6.550  & 7.362  &   7.567  \\
\midrule
\multicolumn{5}{c}{\textbf{LQI}}    \\
 MSE ref.  &  1.954  &   2.544   &    3.399   &    4.957   \\
 MA ene.  &    899    &    861    &  920   &     676   \\
 MA con.  &    1.893  &   3.491   & 5.346      &      5.059   \\
\midrule
\multicolumn{5}{c}{\textbf{PPO2}}    \\
 MSE ref.  &   16.978 &   19.032    &   24.717    &    27.885   \\
 MA ene.  &    531     &   530      &   528   &    526     \\
 MA con.  &   2.525    &     2.718  &    3.230    &  3.546     \\
\midrule
\multicolumn{5}{c}{\textbf{A2C}}    \\
 MSE ref.  &  10.682  &    10.371   &   12.559    &    14.682    \\
 MA ene.  &   732      &    729     &   736   &    731     \\
 MA con.  &  1.608     &   1.536    &   1.690     &  1.883   \\
 \midrule
\multicolumn{5}{c}{\textbf{ACKTR}}    \\
 MSE ref.  &  9.556  &    10.264  &   11.412    &  15.473   \\
 MA ene.  &   1496      &   1524      &   1513   &     1510    \\
 MA con.  &  0.557     &   0.487    &   0.585    &   0.768    \\
\bottomrule
\end{tabular}
    \label{tab:control:GT}
\end{table}

% \paragraph{Discussion}
The  results from Table~\ref{tab:control:GT} demonstrate 
robust performance of the proposed DPC in the reference tracking and constraints satisfaction against all compared methods while paying a cost in terms of increased energy use.
The zero values of the constraints violations (MA con.) demonstrate that the learned DPC policy is capable of $100\%$ constraints satisfaction also in the presence of parametric and additive uncertainties, while simultaneously optimizing the reference tracking objective.
 These results indicate that the proposed DPC policy optimization is a competitive alternative to classical MPC approaches.
An interesting observation is that 
while using the same objective weights as all MPC variants, the proposed DPC method favors the constraints satisfaction and tracking performance before the energy use minimization. This can be explained by the batch optimization of the the control policy that penalizes average control performance over a distribution of samples thus yielding robust behavior across a variety of initial conditions and reference signals.

% classical control
As expected, the performance of the LQR and LQI on the given control task is unsatisfactory, as these classical control approaches are not designed to handle  control problems with inequality constraints.
% RL
The investigated deep RL algorithms are 
able to outperform classical control methods in terms of constraints satisfaction due to the rewards augmented with constraints penalties.
However, all deep RL algorithms perform significantly worse in reference tracking and constraints satisfaction than the proposed DPC policy and the three variants of MPC.
Given greater opportunity to explore the state space beyond the $30,000$ sampled training episodes, the reinforcement learning algorithms could potentially perform better on the provided metrics as deep RL algorithms typically require a large number of samples to converge to a near-optimal solution.
However, DPC policy optimization demonstrates significantly better sampling efficiency against the investigated RL methods with the same number of gradient updates.  

\subsubsection{Adaptive Policy Optimization with Unknown System Model}

In this scenario, the system dynamics matrices of the controlled system~\eqref{eq:truth:model:uncertain} are considered unknown and must be learned by DPC from data. 
For the simultaneous system identification and constrained control learning task  we train the parametrized closed-loop dynamics model as defined in Algorithm~\ref{algo:model_MPCpolicy}.
% with one-step ahead prediction horizon $N = 1$.
% The model is trained with Adam optimizer with learning rate of $0.001$ on $40,000$ epochs.
% The system model and the policy begin training with randomly initialized weights.
% The same train, validation, and test sets are used as  described in section~\ref{sec:results:control}.
% The weights of the loss function~\eqref{eq:SysID_policy_loss} for the control part remain unchanged as given in section~\ref{sec:results:control}, the modeling weights are: $Q_{ID} = 1e2$, $Q_{\Delta x} = 1e0$.
% The results are reported on the test set.
% The simulation model during the closed-loop control 
% is affected by  uncertainties as defined via eq.~\eqref{eq:truth:model:uncertain}.
To demonstrate the adaptive capabilities we apply online updates of DPC policy by optimizing the objective~\eqref{eq:SysID_policy_loss} over $10$ epochs  at each time step during the test set.

Table~\ref{tab:SysID_Control} shows the closed-loop control performance of the adaptive DPC policy
evaluated on  the modeling MSE (MSE mod.), reference tracking MSE (MSE ref.), energy use (MA ene.),  and constraints violation (MA en., MA con.).
% for trained policy with learned model evaluated on $20$ simulation  runs.
Interestingly, the constraint satisfaction performance is comparable to robust MPC designed with the ground truth model, while the reference tracking performance is significantly less conservative.
% The performance of the joint model and policy learning, together with the corrective effects of the adaptive model and updates during online evaluation demonstrate the competitive nature of the proposed methodology compared to classical approaches. 
\begin{table}
    \centering
       \caption{Closed-loop control performance of adaptive DPC on  model prediction, reference tracking,  energy use, and constraints violations  of simultaneous model and  policy learning evaluated on closed-loop simulations with various degrees of uncertainties ${\bf v}_k$ and ${\bf w}_k$.}
\begin{tabular}{lllll}
\toprule
{Test set} &  No unc.  & ${\bf w}_k$  & ${\bf v}_k$  & ${\bf w}_k$ \& ${\bf v}_k$   \\
\midrule
MSE mod.  & 0.344     &    0.485 &   1.325    &   1.781    \\
MSE ref.  & 0.952     &  1.126   &   2.263   &    2.932     \\
MA ene.  & 1102   &  1139     &     1134  &    1109   \\
MA con.  &  0.000    &  0.002  &   0.014  &   0.053    \\
\bottomrule
\end{tabular}
    \label{tab:SysID_Control}
\end{table}
% The non-zero values of MSE for the model and reference tracking are  caused by hard constraints on states and control actions, causing situations when the reference trajectory is not reachable. 
Due to the large weight on the constraints violation penalties, the learned policy deems the state and input constraints satisfaction a primary task.  Hence, the policy may sacrifice reference tracking performance when the constraints could be compromised. 

\begin{figure}
    \centering
    \hspace{-0.3cm}
    \includegraphics[width=.5\textwidth]{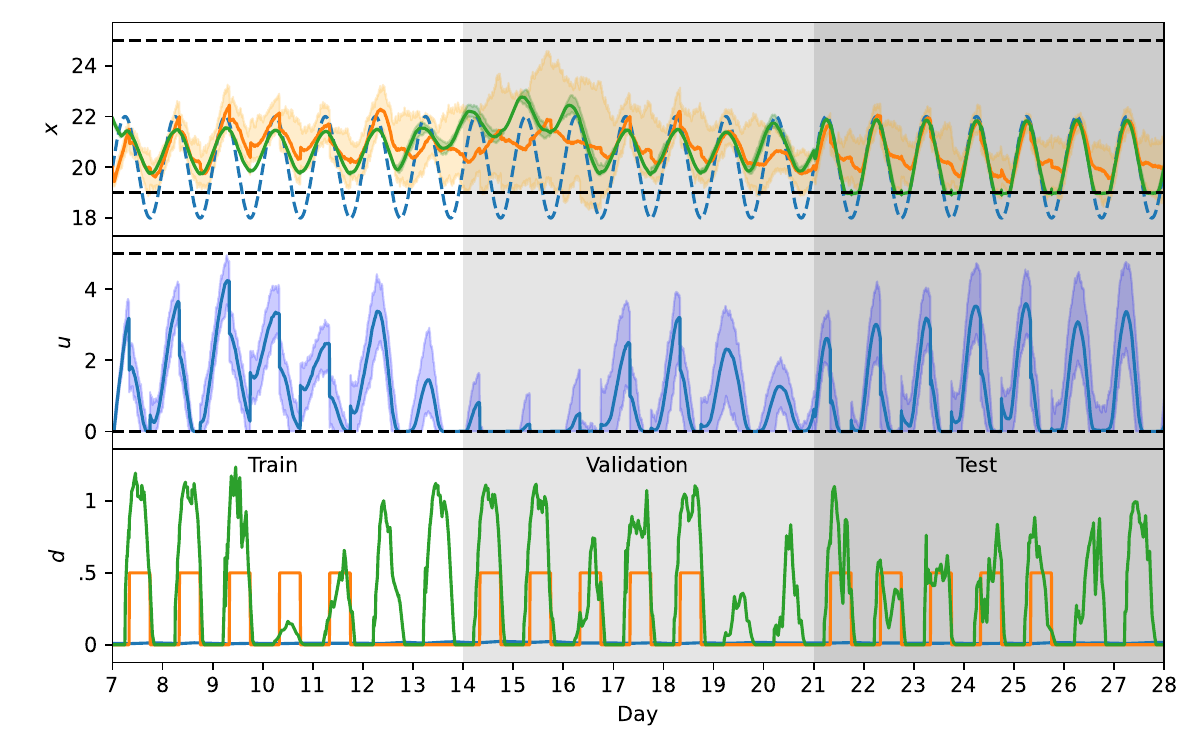}
    \caption{Closed-loop control trajectories  of simultaneous system identification and adaptive DPC policy learning.}
    \label{fig:SysID_Control:cl}
\end{figure}

Figure~\ref{fig:SysID_Control:cl} shows the  closed-loop simulations for the training, validation and test sets with simulation model affected by additive uncertainty ${\bf w}_k$.
The upper plot shows the trajectories of the controlled  (orange) and learned predicted (green) state. The middle plot shows  control's policy actions, while the bottom plot shows the influence of measured disturbances.
Error bars  represent the influence of various realizations of uncertainties. 
These plots demonstrate that adaptive DPC policy can consistently predict the trajectory and track the reference of a controlled state while satisfying the state and input constraints most of the time. 
% Please note that parts of the reference trajectory (dashed blue) were deliberately chosen to be outside the state constraints (dashed black) to demonstrate the balancing capabilities of  the multi-objective loss~\eqref{eq:SysID_policy_loss}.
Online learning  was active only during the test set, resulting in decreased difference between predicted (green) and controlled (orange) state, as well as the variance of the controlled (orange) state.
As a consequence, the constraints handling was also improved with online updates.

 \subsection{Scalability and Performance Guarantees}

In this section, we analyze online (evaluation) and offline (training) computational complexity of the proposed DPC policy optimization method.
The empirical analysis in this section was performed on a laptop with 2.60 GHz Intel(R) i7-8850H CPU and 16 GB RAM on a 64-bit operating system.

\subsubsection{Online computational complexity}
From the online computational and memory standpoint we compare the proposed method with 
implicit  and explicit solutions.
The implicit MPC problem  is solved online via quadratic programming (QP) using Matlab's  Quadprog solver.
In theory, the online complexity of the QP problem depends on the number of constraints, which scale polynomially with the prediction horizon steps $N$, and decision variables. In particular, the theoretical complexity of the QP problem
in a dense form is $\mathcal{O}(N^3n_u^3)$, and in the sparse form is $\mathcal{O}(N^3(n_x+n_u)^3)$~\cite{Frison2013}.
The online procedure of eMPC consists of evaluating the piecewise affine policy~\eqref{eq:pwa_law}. The forward pass of the PWA map is typically performed using binary search that has logarithmic complexity $\mathcal{O}(\text{log}(n_{\mathcal{R}}))$, where  $n_{\mathcal{R}}$ is the number of critical regions of the PWA map that is upper bounded exponentially by the number of constraints $q$ as $n_{\mathcal{R}} \le 2^q$.
In case of DPC, the online computational complexity depends solely on the the number of hidden layers $n_L$ and hidden size $n_z$  of the neural control policy~\eqref{eq:dnn}. 
In particular, the evaluation of a single layer neural network consists of matrix multiplication with worst-case asymptotic run-time $\mathcal{O}(n_z^3)$  followed the elementwise activation function with run-time $\mathcal{O}(n_z)$.
The overall theoretical complexity of the neural policy evaluation then becomes $\mathcal{O}(n_L (n_z^3 + n_z))$, thus being decoupled from the sizes of the optimization problem  $n_x$, $n_u$, $N$. This allows us to design neural policies with a tunable upper bound on worst-case computational time for the problems of arbitrary size.
We empirically compare the mean and maximum online evaluation of the iMPC, eMPC, and DPC policies to demonstrate this advantage.

Furthermore, we compare the memory footprint of the trained neural policy of DPC against PWA policy of eMPC.
The assessment of the memory requirements of the implicit MPC is not straightforward, 
as it depends on the memory footprint of the chosen optimization solver, dependencies, and storage of the problem parameters. In this case, we estimate the optimistic lower bound by assessing only the  memory requirements of the Quadprog solver\footnote{The memory footprint estimate of implicit MPC in Table~\ref{tab:CPU_memory} is based on the standalone Python  implementation of the open-source Quadprog solver: \url{https://pypi.org/project/quadprog/}}.

\begin{table}
    \centering
    %   \caption{Scalability of the proposed deep learning-based MPC (DPC) vs explicit  and implicit  MPC policy with increasing prediction horizon $N$.}
    \caption{Online scalability with increasing prediction horizon $N$. Comparison of mean and worst case online computational time per sample, and memory footprint of the proposed DPC policy against implicit (iMPC) and explicit (eMPC) solutions, evaluated on the model from Example 5.} 
\begin{tabular}{lllllll}
\toprule
{$N$} &   1   &   2  &   4  &   6  &  8  \\
\midrule
& \multicolumn{5}{l}{\textbf{mean CPU time [$1e^{-3}$ s]}}    \\
 DPC   & 0.37 &   0.60 &   \bf{1.03} &  \bf{1.50}  &    \bf{1.92} \\
% adaptive DPC   & 33.2 &  &  &  &    \\
eMPC &   \bf{0.15}  &  \bf{0.45}  &   4.04 &   9.48 &  9.76$e^{3}$ \\
iMPC  &  3.00 &  7.86 & 10.52 & 10.31 &  8.53 \\
\midrule
& \multicolumn{5}{l}{\textbf{max CPU time [$1e^{-3}$ s]}}    \\
 DPC  & 9.94 & 15.93 &   \bf{10.06} &  \bf{11.73} &    \bf{10.19} \\
% adaptive DPC   & 167.5 &  &  &  &    \\
eMPC  &   \bf{6.00} &  \bf{10.00} & 23.00  &  82.00 &  97.49$e^{3}$ \\
iMPC  &  53.00  & 40.00 & 68.00 &  73.00 & 92.00 \\
\midrule
& \multicolumn{5}{l}{\textbf{memory footprint [kB]}}    \\
 DPC &   \bf{9.18}  &   \bf{8.96} &  \bf{11.60} &  \bf{15.80} &   \bf{21.60} \\
% adaptive DPC   & 12.50 &  &  &  &    \\
eMPC   &  48 & 1051 & 7523 & 35176  & 459$e^{3}$ \\
iMPC  
% & \multicolumn{5}{c}{{ $\sim$150}}  
      &  $\sim$150  & $\sim$150 & $\sim$150 & $\sim$150 & $\sim$150  \\
\bottomrule
\end{tabular}
    \label{tab:CPU_memory}
\end{table}

In  Table~\ref{tab:CPU_memory} we report the scalability analysis in terms of mean and maximum online evaluation time, and memory footprint as a function of increasing prediction horizon $N$, evaluated on the problem from Example 5.
The results show that the learned DPC policy is on average $5$ to $10$ times faster than iMPC policy, comparable to eMPC on smaller problems, and orders of magnitude faster than eMPC on larger problems.
Compared with eMPC, the DPC has a significantly smaller memory footprint across all scales. Both CPU time and memory footprint of the DPC policies scale linearly with the problem complexity, while eMPC solutions scale exponentially and are practically infeasible for larger-scale problems.
These findings are supported by the fact that DNNs with \texttt{ReLU} layers are more memory efficient than lookup tables for parametrizing the PWA functions representing eMPC policies~\cite{karg2018efficient}.

\begin{remark}
To account for increasing problem complexity, we train a two layer DPC policy, with progressively increasing number of hidden layers $n_{\text{hidden}} = 10 \, N$.
Both, DPC and iMPC policies share the same objectives, however, due to the exponential complexity growth in the case of eMPC we had to implement the move blocking strategy~\cite{CagEtal2007} limiting the length of the control horizon for prediction horizons larger than $4$ as follows: if $N=4$ then $N_c =2$,  if $N=6$ then $N_c =1$, while the problem with $N=8$ is practically intractable even with $N_c =1$.
\end{remark}

\begin{remark}
The online complexity of DPC depends entirely on the choice of the policy parametrizations, i.e., number of layers, and hidden neurons. Therefore it can represent a tunable trade-off between the performance and guaranteed complexity, a design trait desirable for embedded applications with limited computational resources. 
\end{remark}

\subsubsection{Offline computational complexity}

 For both, DPC and approximate MPC  we evaluate the offline computational complexity using a problem from Example 2.
 The resulting scalability analysis of the policy optimization algorithms with the increasing number of training samples is given Table~\ref{tab:dpc_vs_ampc_scalability}.
  In particular, we evaluate total offline computational time as a sum of dataset generation and training time, respectively.
  We demonstrate that the proposed DPC algorithm requires fewer computational resources than aMPC based on imitation learning.
  The first reason is that DPC dataset generation does not require the solution of the original MPC as in the case of aMPC. 
  Interestingly, using the same number of epochs, same policy architecture, and the same learning algorithm hyperparameters, the DPC approach has shown to be faster in the training time than the aMPC supervised learning method.
As observed in the presented example, the policy training via the DPC method seems to be approximately three times faster than aMPC. However, to validate the generality of this statement, one would need to provide rigorous computational analysis of the corresponding learning problems as well as large-scale computational comparison studies that are beyond the scope of this paper.
\begin{table}[ht!]
    \centering
    %   \caption{Scalability of the proposed deep learning-based MPC (DPC) vs explicit  and implicit  MPC policy with increasing prediction horizon $N$.}
    \caption{Offline scalability with the number of training samples. Comparison of the combined dataset generation  and training time of the approximate MPC (aMPC) and the proposed differentiable predictive control (DPC) policy optimization algorithms, evaluated on the problem from Example 2 with prediction horizon $N=10$. } 
\begin{tabular}{lllllll}
\toprule
{\# samples} &   $1e^{3}$  &   $2e^{3}$  &   $3e^{3}$  &   $4e^{3}$  &  $5e^{3}$  \\
\midrule
& \multicolumn{5}{l}{\textbf{dataset generation time [s]}}    \\
 DPC   &  $0$ & $0$   &  $0$ & $0$ &  $0$   \\
aMPC  & $25.2$ &  $50.4$  &  $78.2$ &  $102.0$ &  $130.3$ \\
\midrule
& \multicolumn{5}{l}{\textbf{training time [s]}}    \\
 DPC   &  $57.5$ &  $159.4$  & $241.4$  & $501.8$ &  $630.8$   \\
aMPC  & $222.9$ &  $482.8$  & $717.3$  & $ 871.9$ & $1797.7$ \\
\midrule
& \multicolumn{5}{l}{\textbf{total offline time [s]}}    \\
 DPC   &  $57.5$ &  $159.4$  & $241.4$  & $501.8$ &  $630.8$   \\
aMPC  & $248.1$ &  $533.2$  & $795.5$  &  $973.9$ & $ 1928.0$ \\
\bottomrule
\end{tabular}
    \label{tab:dpc_vs_ampc_scalability}
\end{table}

In this paper, the explicit MPC (eMPC) problems are solved offline via multiparametric quadratic programming (mpQP) using MPT3 toolbox~\cite{MPT3:2013}. Unfortunately, the computational complexity  of the parametric programming problems using analytical solvers scales exponentially with the number of
constraints. In the  multiparametric QP the worst case is given by the all possible combination of active constraints $\mathcal(O)(2^q)$, where $q$ is the number of constraints~\cite{Alessio2009,Bemporad2006}.
As a consequence applications of the eMPC approach are limited to very small problems as shown in Table~\ref{tab:CPU_memory}, where the mpQP problem for larger prediction horizons took several hours to compute.
Thus we omit the eMPC approach from the scalability comparison with DPC and aMPC on a larger problem used for the analysis in Table~\ref{tab:dpc_vs_ampc_scalability}.

%   \add{Aaron: any idea why is aMPC based on supervised learning slower in training than DPC?}

% \add{
% comment on theoretical big O complexity of solving QP - for sampling aMPC datasets \\
% argument: DPC scales linearly with number of samples (Table 4) and number of decision variables (Table 3), aMPC based on online MPC solvers scales polynomialy (TODO evaluate rmpirically), iMPC scales polynomialy (theoretical big O), eMPC scales exponentially  (theoretical big O)
% }

 \subsubsection{Probabilistic performance guarantees}
 
 For both, DPC and approximate MPC  we evaluate the empirical risk $\mu$
 of the trained policies given the formula~\eqref{eq:probability_mu} using a problem from Example 2.
Given a risk tolerance $\epsilon=0.02$ and confidence $0.9835$
 we generate $7000$ closed-loop trajectories by simulating the system~\eqref{eq:closed_loop} with trained policies on samples of normally i.i.d.  initial conditions. 
 Then we calculate the lower bound on empirical risks $\mu$ of stability and constraints violations as a function
 number of training samples and we report the results in Table~\ref{tab:prob_performance}.
The results for DPC and aMPC policies trained on the same amount of data are almost identical,
 where DPC seems to perform slightly better with more training samples, and aMPC does better with fewer training samples.
\begin{table}[ht!]
    \centering
    \caption{Probabilistic performance comparison of the proposed DPC and approximate MPC (aMPC) evaluated using empirical risks $\mu$~\eqref{eq:probability_mu} as a function of training samples. } 
\begin{tabular}{lllllll}
\toprule
{\# samples} &   $1e^{3}$  &   $2e^{3}$  &   $3e^{3}$  &   $4e^{3}$  &  $5e^{3}$  \\
\midrule
 DPC   &  0.9590 &  \textbf{0.9789}  & \textbf{0.9793}  & \textbf{0.9785} &   \textbf{0.9791} \\
aMPC  & \textbf{0.9775}  & 0.9781 & 0.9776 & 0.9742 &  0.9770  \\
\bottomrule
\end{tabular}
    \label{tab:prob_performance}
\end{table}

% \add{add remark on plant model mismatch and future work}

\section{Limitations and Future Work}

The proposed paper presents a proof of concept of the proposed DPC policy optimization method.
One of the main limitations of this study is the assumption of the linear dynamics of the controlled system. The extension to non-linear systems will be addressed in future work.
The presented problem formulation with penalty constraints is an approximation of the original constrained optimization problem. Hence, using the stochastic gradient descent on the proposed parametric loss function does not guarantee the constraint's satisfaction. As a consequence, we need to rely on probabilistic performance guarantees based on the sampled system rollouts. In the future, we will explore using more advanced constrained optimization solvers such as the augmented Lagrangian method.
The probabilistic performance guarantees in this paper are derived based on the assumption of no plant-model mismatch. We will address the extension of the proposed DPC policy optimization algorithm to robust, stochastic, and offset-free variants in future work.
Like MPC, the  DPC policy used in this paper is a full state feedback controller. Therefore a state estimator is necessary for practical applications with partially observable systems. Fortunately, the learned model can be used to design the Kalman filter or moving horizon estimator. An alternative approach is to learn an autoregressive state-space model based on the output measurements~\cite{skomski2021constrained,drgona2020differentiable} as part of the DPC policy. 

\section{Conclusions}

% opens new avenues for research at intersection of ML and controls \\
% stability guarantees and more}

In this work we present a novel direct policy optimization algorithm based on differentiable programming. 
We systematically combine model-based and learning-based principles in the proposed differentiable predictive control (DPC)
method. 
In particular, we provide a new perspective on the use of  differentiable programming for obtaining policy gradients that can lead to solutions of the parametric programming problems arising in explicit model predictive control (MPC).
The presented DPC policy optimization method can learn stable neural control policies subject to state and input constraints.  
The neural control policy, as well as the system dynamics model, can be learned end-to-end without the need for an expert policy to imitate. 
We provide an algorithm for obtaining stochastic performance guarantees of the trained policies for safety verification.

We demonstrate the performance in the extensive case studies, evaluating the control performance, data efficiency, and scalability of the proposed DPC method  against classical controllers, implicit and explicit MPC, approximate MPC,
and model-free deep reinforcement learning (RL) algorithms.
We demonstrate that DPC can learn stabilizing neural control policies for unstable linear systems and adaptive control policies for unknown systems while requiring less data and fewer policy updates than model-free RL algorithms. Moreover, we show that DPC is faster than implicit MPC, has a smaller memory footprint than explicit MPC, and requires fewer resources for training than approximate MPC.

Based on the reported results and user-friendly open-source implementation~\cite{Neuromancer2021}, the proposed DPC has the potential to be deployed in the application domains beyond the computational reach of the traditional explicit MPC while providing performance guarantees which are typically lacking in purely data-driven approaches.
Moreover, we believe that due to the appealing theoretical connections with MPC, the proposed methodology provides new research opportunities on a systematic combination of modern data-driven methods with mature control-theoretic concepts.

\section*{Acknowledgements}
We would like to thank Jackson Warley and Wenceslao Shaw-Cortez for fruitful discussions that helped improve the technical quality of the presented ideas.
Also, we want to thank our anonymous reviewers for their constructive feedback.

This research was supported by 
the Mathematics for Artificial Reasoning in Science (MARS) initiative
via the Laboratory Directed Research and Development (LDRD) investments at Pacific Northwest National Laboratory (PNNL). PNNL is a multi-program national laboratory
operated for the U.S. Department of Energy (DOE) by Battelle Memorial Institute under Contract
No. DE-AC05-76RL0-1830.

\bibliographystyle{IEEEtran}
\bibliography{main}

\section*{Appendix}

\subsection{Mathematical Background}
\label{sec:math_bck}

\begin{definition}
{Induced operator norm} of $\mathbf{A} \in \mathbb{R}^{n\times m}$ is given as:
 \begin{equation}
 ||\mathbf{A}||_p =  \max_{\mathbf{x} \neq 0} \frac{ ||\mathbf{A}\mathbf{x}||_p}{||\mathbf{x}||_p}  =   \max_{\|\mathbf{x}\|_p = 1} ||\mathbf{A}\mathbf{x}||_p, \ \ \forall \mathbf{x} \in \mathcal{X},
   \label{eq:operator_norm}
 \end{equation}
%  \begin{equation}
%  ||\mathbf{A} \mathbf{x}|| \leq c ||\mathbf{x}||, \ \ \forall \mathbf{x} \in \mathcal{X}
%   \label{eq:operator_norm}
%  \end{equation}
 where $\mathcal{X}$ is a normed vector space, and  $|| \cdot ||_p: \mathbb{R}^{n} \to \mathbb{R}$ represents the vector $p$-norm inducing the operator norm $ || \mathbf{A} ||_p: \mathbb{R}^{n \times m} \to \mathbb{R}$.  
%  Then the  operator norm of a matrix $\mathbf{A}$ gives an upper bound of its spectral radius:
%  \begin{equation}
%   \label{eq:operator_norm_bound}
% \rho(\mathbf{A}) \le || \mathbf{A} ||_p.
% \end{equation} 
 \end{definition}
 \begin{definition}
{Induced $p$-norm} $||\cdot||_p: \mathbb{R}^{n \times m} \to \mathbb{R}$ is called submultiplicative if it satisfies~\cite{matrix_norms1983}:
 \begin{equation}
  \label{eq:operator_norm_submultiplicative}
|| \mathbf{A} \mathbf{B} ||_p \le || \mathbf{A}||_p  ||\mathbf{B} ||_p.
\end{equation} 
\end{definition}
\begin{theorem}~\cite{LinAlg2020}
Lets have a vector norm $||\cdot||_p: \mathbb{R}^{n} \to \mathbb{R}$ defined for all $n$ with corresponding induced operator norm defined as~\eqref{eq:operator_norm},
 then the submultiplicatity of the operator norm~\eqref{eq:operator_norm_submultiplicative} is satisfied for any $\mathbf{A} \in \mathbb{R}^{m \times k}$ and
  $\mathbf{B} \in \mathbb{R}^{k \times n}$.
  \label{thm:submult}
\end{theorem}

\begin{definition}
Given a metric space $(\mathcal{X},d)$, a mapping $f: \mathcal{X}\to \mathcal{X}$ is called contractive if there exist a metric $d$, a constant $c \in [0, 1)$, and the following holds:
 \begin{equation}
  \label{eq:contraction}
      d(f( \mathbf{x_1}),f( \mathbf{x_2})) \le  c d( \mathbf{x_1}, \mathbf{x_2}), \  \forall  \mathbf{x_1},  \mathbf{x_2} \in   \mathcal{X}
\end{equation} 
\end{definition}
% https://en.wikipedia.org/wiki/Banach_fixed-point_theorem

\begin{definition}
\label{def:affine_contract}
An affine map $f(\mathbf{x}) =  \mathbf{A} \mathbf{x} + \mathbf{b}$ with $\mathbf{x} \in \mathbb{R}^n$ and  metric $d = ||\cdot||_p$ is contractive if  $ ||\mathbf{A}||_2 < 1$.
\end{definition}
% https://sun4.vaniercollege.qc.ca/~iti/proj/David.pdf

\begin{theorem}
\label{thm:banch}
Banach fixed-point theorem. In a non-empty complete metric space $(\mathcal{X},d)$ every contractive map~\eqref{eq:contraction} is converging towards a unique steady state (fixed point)  ${\bf x}_{ss}  =  
   \lim_{t \to \infty} f({\mathbf{x}}_t)$.
\end{theorem}

\begin{lemma}~\cite{Hertneck8371312,vonluxburg2008statistical}
\label{lema:Hoeffding}
Let $I(\mathbf{X}^i)$ for $i \in \mathbb{N}_1^m$ be iid random variables. Then  the following inequality holds
\begin{equation}
    \mathbb{P}[|\mu  - \tilde{\mu }| \ge \epsilon ] \le 2 \text{exp} (-2 m  \epsilon^2)
\end{equation}
with the right hand side representing the confidence level $\delta := 2 \text{exp} (-2 m  \epsilon^2)$.
Where $m$ gives number of samples, $\mu$ is the risk factor,
$\tilde{\mu }$ stands for the empirical risk,
and $\epsilon$ 
represents the risk tolerance.
Then it holds that with confidence at least $1-\delta $ we have:
\begin{equation}
\label{eq:probability_mu}
    \mathbb{P}[I(\mathbf{X}^i)=1] = \mu  \ge \tilde{\mu } -  \epsilon
\end{equation}
\end{lemma}

% \subsection{System Dynamics Models from Examples 2, 3, 4}
% \label{sec:results:models}

% \add{add SSM parameters}

% \subsubsection{PVTOL aircraft model}

% \subsubsection{Building thermal dynamics model}

\subsection{Approximate Model Predictive Control}
\label{sec:results:control}
In example 2, we compared the performance of the proposed DPC policy optimization against approximate MPC (aMPC) method based on imitation learning of the original implicit MPC solutions.
The aMPC  is given as a supervised learning problem with the following loss function:
\begin{equation}
  \mathcal{L}_{\texttt{aMPC}} =  \min_{{\bf W}}  \sum_{i=1}^{m}   || {\bf u}_{\texttt{MPC}}^i - {\bf u}_{\texttt{aMPC}}^i ||_2^2  
\end{equation}
where ${\bf u}_{\texttt{MPC}}^i$ are supervisory signals of control actions computed by solving implicit MPC problem~\eqref{eq:mpc_example} via constrained optimization solver. 
The aMPC policy is given by neural network~\eqref{eq:dnn} that is mapping a vector of control parameters $\boldsymbol\xi^i$ onto control cations ${\bf u}_{\texttt{aMPC}}^i$, i.e.:
\begin{equation}
{\bf u}_{\texttt{aMPC}}^i = \boldsymbol \pi^{\texttt{aMPC}}_{\bf W}(\boldsymbol\xi^i)
\end{equation}
The $i$-th index here represents a sample from the distribution of the control parameters $\boldsymbol\Xi$ used to generate the dataset via the solutions of a single implicit MPC problem per each sample.

% \add{add approximate MPC method in the appendix for comparison with all the hyperparameters. Highlight differences with imitation learning and our approach. }

\subsection{Hyperparameters of Control Methods from Example 5}
\label{sec:results:control}

\subsubsection{Differentiable predictive control}
For continuous control task  we train
the proposed DPC policy  as defined in algorithm~\ref{algo:DPC_optim} by optimizing the loss function~\eqref{eq:policy_loss}
with one-step ahead prediction horizon $N = 1$.
  The choice of the \del{layer of} neural control policy ${\bf u}_k =  \pi_{{\bf W}}({\boldsymbol \xi}) $  depends on the problem complexity. 
  For the purposes of this paper we used a two-layer neural network  $\pi_{{\bf W}}({\boldsymbol \xi}) : \mathbb{R}^{n_{\boldsymbol \xi}} \to \mathbb{R}^{N n_u} $ with \texttt{ReLU} activations,  and policy features $\boldsymbol \xi = \{{\bf x}_k, {\bf r}_k, \underline{{\bf x}}_k, \overline{{\bf x}}_k, \underline{{\bf u}}_k, \overline{{\bf u}}_k \}$.
  The number of hidden units in each layer given by $n_{\text{hidden}} = 10 \, N$.
The policy is trained with Adam optimizer \cite{kingma2014adam} and learning rate of $0.001$ on $30,000$ epochs.
The  policy begins training with randomly initialized weights.

For learning the DPC policy~\eqref{eq:policy_loss}, the trajectory  $\mathbf{R}$ is uniformly sampled in the realistic operative range of \SIrange{15}{25}{\celsius} to increase the generalization.
   The simulated  state trajectories, $\mathbf{X}$,
 are 
 uniformly sampled state distribution in realistic range  \SIrange{0}{25}{\celsius}  to robustify the learned DPC  policy.
 The  control inputs trajectory $\mathbf{U}$  is  generated by the control policies.

The different one week periods of $2016$ samples are used as training, validation, and test sets, respectively.
The weights of the loss function~\eqref{eq:policy_loss} are tuned using the physical insight about the system dynamics  with the following values:
$Q_r = 2e1$, 
$Q_u = 1e-6$, $ Q_g = 5e-7$, $Q_h  = 5e1$.
The results are reported on the test set.

% comparison with classical control 
\subsubsection{Classical control methods}
We compare the performance of the trained DPC policy  with classical model-based linear control methods, namely
linear quadratic regulator (LQR), and  Linear quadratic integral controller (LQI) implemented in Matlab environment using commands \texttt{dlqr} and \texttt{lqi}, respectively.
The weights of LQI controller $Q_r$, and $Q_u$ are chosen to be identical to the DPC policy, while tuning of the LQR weights was necessary to obtain satisfactory performance.

% MPC
\subsubsection{Model predictive control} We evaluate the performance of three different formulations of model predictive control (MPC) problem. 
The nominal MPC is obtained by the solution of the constrained optimization as defined by~\eqref{eq:mpc_example} with softened input and state constraints. 
Robust MPC (RMPC) represents the worst-case scenario approach via constraints tightening techniques~\cite{SR00,Lofberg_minimaxRMPC2003}. 
On the other hand, stochastic MPC (SMPC)
is based on probabilistic constraints approximated by sampling from the known distribution of the unknown disturbances. 
The implementation of RMPC and SMPC is based on the formulations presented in~\cite{Drgona:CDC13}.
% hyperparameters
For a fair comparison, all MPC problems are designed with identical setup of the hyperparameters ($N$, $Q_r$, $Q_u$, $Q_h$, and $Q_g$) as in the case of trained DPC policy.
% MPC solutions
The MPC problems are implemented in the Matlab environment using the Yalmip optimization toolbox~\cite{yalmip:2004}.

% deep RL
\subsubsection{Deep reinforcement learning}
Additionally, we assess the performance of three deep reinforcement learning (RL) algorithms suitable for dealing with continuous state and action spaces obtained from the Stable Baselines implementation~\cite{stable-baselines2018}.
In particular, we investigate an asynchronous, deterministic variant of  Advantage Actor Critic (A2C)~\cite{MnihBMGLHSK16},
Actor Critic using Kronecker-Factored Trust Region (ACKTR)~\cite{ACKTR2017}, and  Proximal Policy Optimization algorithm~\cite{SchulmanWDRK17}. We design the reward to be equal to the negative value of the MPC loss~\eqref{eq:policy_loss}  using the same values of the weights $Q_r = 2e1$, 
$Q_u = 1e-6$, $ Q_g = 5e-7$, $Q_h  = 5e1$.
For a fair comparison, we train randomly initialized RL policies on $30,000$ episodes over the same number of time steps as DPC with a learning rate of $0.01$.
We use the ground truth system dynamics model as the training environment.

\end{document}